%% file: p3.tex
\documentclass[12pt]{article}
\usepackage{epsfig}

\renewcommand{\Re}{{\rm Re}}
\renewcommand{\Im}{{\rm Im}}

\newcommand{\A}{{\bf A}}
\newcommand{\D}{{\bf D}}
\renewcommand{\j}{{\bf j}}
\renewcommand{\k}{{\bf k}}
\newcommand{\kr}{{\bf k}\cdot {\bf r}}
\newcommand{\tk}{\widetilde{\bf k}}
\newcommand{\tkx}{\tilde{k}_x}
\newcommand{\tky}{\tilde{k}_y}
\newcommand{\tq}{\tilde{\bf q}}

\newcommand{\q}{{\bf q}}
\renewcommand{\r}{{\bf r}}
\newcommand{\rp}{{{\bf r}'}}

\newcommand{\Ref}[1]{\ref{#1}}
\renewcommand{\S}{{\bf S}}
\newcommand{\FFXY}{FF$XY$}
\newcommand{\Lmin}{L_{\rm min}}
\newcommand{\m}{{\bf m}}
\newcommand{\M}{{\bf M}}

\newcommand{\TKT}{T_{\rm KT}}
\newcommand{\TKTL}{T_{\rm KT}^{(L)}}

\newcommand{\TCG}{T^{\rm CG}}
\newcommand{\parti}[2]{\frac{\partial #1}{\partial #2}}
\newcommand{\vsigma}{{\bf \sigma}}
\newcommand{\muhat}{{\hat{\bf \mu}}}
\newcommand{\xhat}{{\hat{\bf x}}}
\newcommand{\yhat}{{\hat{\bf y}}}
\newcommand{\onlinecite}[1]{\cite{#1}}

\newcommand{\xlabel}[2]{\put(#1,-1){\makebox(0,0)[t]{#2}}}
\newcommand{\ylabel}[2]{\put(-1,#1){\makebox(0,0)[r]{#2}}}
\newcommand{\rotate}[1]{
   #1 
}

\begin{document}
\title{Monte Carlo study of the Villain version of the fully frustrated
  $XY$ model} 
\author{Peter Olsson}
\date{\today}
\maketitle
\begin{abstract}
  The fully frustrated $XY$ model with Villain interaction on a square
  lattice is studied by means of Monte Carlo simulations.  On the basis of
  the universal jump condition it is argued that there are
  two distinct transitions in the model, corresponding to the loss of $XY$
  order and $Z_2$ order, respectively.  The Kosterlitz-Thouless (KT)
  transition is analyzed by finite size scaling of the helicity modulus at
  lattices of size $L = 32$ through 128, giving $T_{\rm KT} \approx
  0.8108(1)$.

  The vorticity-vorticity correlation function is used to determine
  two different characteristic lengths, the $Z_2$ correlation length
  $\xi$, and the screening length $\lambda$, associated with the KT
  transition and free vortices. The temperature dependence of $\xi$
  is examined in order to determine $T_c$ and the correlation length
  exponent, $\nu$.  The exponent is found to be consistent with the 2D
  Ising value, $\nu = 1$, and the obtained critical temperature is
  $T_c = 0.8225(5)$.  The determinations of both $\xi$ and $\nu$ are
  done carefully, first applying the techniques to the 2D Ising
  model, which serves as a convenient testing ground.
\end{abstract}

\newpage

\section{Introduction}

The critical behavior of the fully frustrated $XY$ (\FFXY) model has
received much attention during the last decade. This is due to the two
kinds of symmetries present in the systems and the associated possibility
of new critical behavior. But in spite of the large number of papers, there
is still no consensus about the phase transition(s) in this model.

The ordinary frustrated $XY$ model (with cosines interaction) is governed
by the Hamiltonian
\begin{displaymath}
  H = -J \sum_{\left<ij\right>} \cos(\theta_i - \theta_j - A_{ij}),
\end{displaymath}
where $i$ and $j$ enumerate the lattice sites, $\theta_i$ is the angle at
lattice point $i$, $A_{ij}$ is the quenched vector potential, and the sum is
over nearest neighbors. The frustration is determined by the sum of
$A_{ij}$ around a plaquette (see below). In the fully frustrated case --
examined in the present paper -- this sum is equal to $\pi$.

The peculiarities of the fully frustrated models stem from the two
different symmetries. Beside the rotational symmetry of the $XY$ model, the
model also has an $Z_2$ symmetry associated with the chirality. The
square-lattice version in the ground state has a checkerboard pattern of
plaquettes with positive or negative chirality, corresponding to clockwise
or counterclockwise rotation of the spins\cite{Villain:77}. This is the
same symmetry as in the anti-ferromagnetic Ising model.

Some other realizations of \FFXY\ models are the anti-ferromagnetic $XY$
model on a triangular lattice\cite{Miyashita_Shiba}, the Coulomb gas with
half-integer charges\cite{Thijssen_Knops:88}, and the 19-vertex version of
the \FFXY\ model\cite{Knops_NKB}. All these models are generally assumed to
have similar critical behavior.  This is also expected to be true for the
$XY$ Ising model\cite{Granato_KLN} for certain choices of some parameters.

In the first MC simulations\cite{Teitel_Jayaprakash:83a} Teitel and
Jayaprakash found a divergence in the specific heat, consistent with an Ising
transition, accompanied by a steep drop in the helicity modulus. But since the
data did not allow for any precise determination of the critical
temperature(s), the authors suggested two possible scenarios:
\begin{enumerate}
\item At temperatures closely below $T_c$, the Ising excitations give
  rise to a steep drop in $\Upsilon$. When $\Upsilon$ approaches
  $2T/\pi$, the vortex excitations take over and produce an universal jump
  at a temperature $\TKT < T_c$. This means two distinct transitions in
  well-known universality classes. 
\item The Ising excitations give a transition with an associated
  non-universal jump at the same temperature as the peak of the
  specific heat. This alternative is a single transition in a new
  universality class.
\end{enumerate}
While these often have been considered the main options, several other
possibilities have also been suggested in the literature, as e.g.\ an Ising
transition at a lower temperature than the Kosterlitz-Thouless (KT)
transition.\cite{Garel_Doniach}

Over the years there have appeared several reports of MC studies where the
losses of both $XY$ and $Z_2$ orders have been studied. For the
anti-ferromagnetic $XY$ model on a triangular
lattice\cite{Miyashita_Shiba}, a study of the heat capacity and the $XY$
susceptibility suggested two distinct transitions, with $\TKT < T_c$. The
temperature difference was, however, quite small and the possibility of a
single transition could not be ruled out. In a second study of the same
model\cite{LJN_Landau}, including somewhat larger lattices and with a
careful analysis of the $Z_2$ transition, the transitions were found to be
even closer together. The temperature difference was in this study well
below the statistical uncertainty.  The results were therefore suggestive
of a single critical point.

Likewise, there are conflicting results for the Coulomb gas with
half-integer charges in the literature. The first study, by Thijssen and
Knops\cite{Thijssen_Knops:88}, suggested coinciding transitions whereas
Grest found two distinct transitions\cite{Grest:89}.  The conflicting
values of the temperature for the loss of $XY$ order was apparently due to
different methods to locate the transition. In the first case $\TKT$ was
determined from the maximum finite size dependence in $1/\epsilon$, and in
the second case from the crossing of $1/\epsilon$ for different system
sizes.  The latter method gives a lower value of the transition
temperature. The different results for the $Z_2$ transition seem to be due
to differences in the MC data.  Whereas the earlier study reported a drift
in position of the peak in the heat capacity to lower temperatures with
increasing lattice size, such a size dependence was not verified in the
latter simulation. This discrepancy gave higher values of $T_c$.  Similar
conclusions were also obtained from recent simulations of the Coulomb gas
with half-integer charges\cite{J-R.Lee:94}. The results from this study
were two distinct transitions; at a lower temperature a KT transition with
a non-universal jump, followed by a $Z_2$ transition with non-Ising
exponents (see below).

As discussed in Ref.\ \cite{Olsson:self-cons.long} the dielectric constant
at smallest possible wave vector $k = 2\pi/L$ from the CG simulations, is
not an ideal quantity for locating the KT transition. With this kind of
boundary conditions, there are two finite-size effects working in the
opposite directions. That this quantity is more or less size-independent
only means that these two effects happen to nearly cancel each other. This
casts doubt on both the KT temperatures and the non-universal jumps found
in the above studies.

In order to circumvent the difficulties associated with a precise
determination of $\TKT$ it has been argued that a determination of the
critical exponents for the $Z_2$ transition by means of finite size
scaling, would be the best way to arrive at some firm conclusions.
This, at first, seems as a good idea since the study of finite size
effects right at $T_c$ usually is the by far most efficient way to
extract the critical behavior by MC simulations.

In this spirit the correlation length exponent $\nu$, has been determined
in a fairly large number of studies by means of finite size scaling at
$T_c$. In the MC simulations
\cite{L_Kosterlitz_G,Lee_Lee,Ramirez-Santiago_Jose:94} this exponent is
extracted from the temperature dependence of various kinds of measures of
the distribution of the staggered magnetization.  The same exponent has
also been obtained from transfer matrix
calculations\cite{Granato_Nightingale}.  The results are generally in favor
of non-Ising exponents, $\nu = 0.85(3)$, 0.813(5), 0.875(35), 0.80(4), and
the critical temperatures $T_c/J = 0.455(2)$, 0.454(2), 0.4206, 0.454(4).
Determinations of the same exponent in the 19-vertex version of the \FFXY\ 
model\cite{Knops_NKB}, the Coulomb gas with half-integer
charges\cite{J-R.Lee:94}, and the $XY$ Ising
model\cite{Granato_KLN,Nightingale_GK}, gave $\nu = 0.77(3)$, 0.84(3),
0.85(3), and 0.79, respectively.  It seems, however, to be the case that
such finite size scalings in many cases are not quite satisfactory and
therefore not conclusive\cite{Granato_KLN,Granato_Nightingale,J-R.Lee:94}.

With the steadily increasing computational resources it has been possible
to obtain data with high precision for increasingly larger lattices. A
recent paper reported results for the helicity modulus at a $L = 128$
system for the first time\cite{Lee_Lee}. These data has far-reaching
implications since it was shown that the helicity modulus crosses the
universal line, $2T/\pi$, at a surprisingly low temperature, well below the
temperatures quoted above for the $Z_2$ transition.  This must be
considered very strong evidence that the $XY$ order is lost at a
temperature below the $T_c$ obtained from finite size scaling; and thus
exclude the single transition scenario.  However, with this position the
non-Ising exponents become problematic. The non-Ising exponents are usually
explained as an effect of the interaction between $XY$ and Ising critical
excitations -- a reasonable explanation only if the two kinds of order are
lost at the same temperature.

We therefore have two pieces of evidence pointing in opposite directions.
The presence of non-Ising exponents strongly suggests a single transition,
whereas the early drop in the helicity modulus seems to exclude this
possibility.

A consistent view of these matters was recently suggested in Ref.\ 
\cite{Olsson:xyff}. The key observation is that a consequence of a KT
transition below $T_c$ would be the presence of a finite but large
screening length $\lambda$, at the $Z_2$ transition temperature, $T_c$.
For finite size scaling to be valid, it is necessary that $L$ be much
larger than all other finite length scales in the system, and in particular
$\lambda$.  The large value of $\lambda$ could therefore invalidate earlier
finite size scaling analyses.  The condition for a successful application
of finite size scaling at $T_c$, $L\gg \lambda$, may imply very large
systems.

In the most ambitious study so far of the $XY$ Ising model by means of
Monte Carlo transfer matrix calculations on infinite strips with widths up
to 30 lattice spacings, Nightingale \emph{et al.} again found evidence for
non-Ising exponents\cite{Nightingale_GK}. They did, however, also find an
`internal inconsistency' in two different determinations of the thermal
exponent $y_T$, which led them to call in question the applicability of
scaling theory. This inconsistency is certainly in line with the suggested
failure of finite size scaling due to the finite screening length
$\lambda$.

The main results in Ref.\ \cite{Olsson:xyff} were a precise determination
of $\TKT$, together with a demonstration that the staggered magnetization
is, indeed, influenced by the screening length $\lambda$, unless $L \gg
\lambda$, i.e.\ the helicity modulus $\Upsilon \approx 0$.  In order to
show that the behavior is consistent with the Ising exponent $\nu = 1$, the
behavior of the correlation length was also examined.  This part of the
study was, however, hampered by two different complications. First, the
correlation function did only fit nicely to an exponential decay for
temperatures pretty far away from $T_c$. Second, it was difficult to
include the effect of the spin waves in an entirely convincing manner.
While it is certainly possible to argue in favor of the employed
technique\cite{Olsson:xyff}, this is at best only an approximative way to
compensate for the temperature-dependent effects of the spin waves.

One of the aims of the present study is to improve on the problematic
points in the temperature dependence of the correlation length. The
complications with temperature-dependent effects of the spin waves is taken
care of by performing simulations in an \FFXY\ model with Villain
interaction -- the model dual to the CG with half-integer charges. The
point is that both the vortex interaction and the vorticity ($\pm 1/2$) are
manifestly temperature-independent in that model.  To find a reliable
technique for determinations of the correlation length, we compare with the
behavior in the 2D Ising model. In that case we benefit from the dual
advantages of a fast cluster algorithm and exact knowledge of the critical
behavior. From the simulations of the 2D Ising model we show that the
region where the true Ising exponent may be found is very narrow.  A similar
analysis of the \FFXY\ model gives the same kind of conclusion, and it
therefore seems that the data points in Fig.\ 5 of Ref.\ \cite{Olsson:xyff}
actually are outside the critical region.

The main result of the present analysis of the fully frustrated 2D $XY$
model with Villain interaction is the existence of two distinct
transitions. An ordinary Kosterlitz-Thouless transition at $\TKT /J =
0.8108(1)$ followed by an Ising transition at $T_c/J = 0.8225(5)$, about
1.4 \% above.  This is a fairly small temperature difference but, as we
will see below, the conclusion of two distinct transitions is not built on
an estimate of the temperature difference between two separate transitions.
Section \Ref{Sec:Finite.univ.two} gives a strong argument for the existence
of two transitions which is not based on the determinations of the two
transition temperatures.

The organization of the present paper is as follows: In Sec.\ II we define
the model, describe the quantities measured in the simulations and some of
the analyses to be performed on the data. Section III begins our analyses
of MC data. We shortly describe the MC procedure employed to obtain the
data and some checks used to validate the results.  The major part of
Section III gives the results from various analyses that take advantage of
the finite size dependence in the MC data.  Among these are the new
argument for two distinct transitions, the determination of $\TKT$ through
finite size scaling of the helicity modulus, and an analysis of Binder's
cumulant for the staggered magnetization.

Section IV contains the determinations of the characteristic lengths $\xi$
and $\lambda$ from the correlation function.  In this paper $\xi$ denotes
the correlation length associated with the Ising-like degrees of freedom,
whereas $\lambda$ is the screening length associated with the KT transition,
which (besides a constant factor) is equivalent to the $XY$ correlation
length. Since the finite size effects in this context are unwanted
complications, we take some pains to examine the appearance of finite size
effects.  In order to test some techniques for the analysis of correlation
functions, and the critical behavior from the correlation length, we make
use of the 2D Ising model as a testing ground.  After these preliminaries
we employ these techniques to the correlation functions from the \FFXY\ 
model to determine the temperature dependence of both $\xi$ and $\lambda$
above $T_c$, and $\xi$ at low temperatures.  This section also contains an
examination of the effect of domain walls on the vortex interaction.

Finally, in Section V we put our results in relation to some results by
Berg\'{e} et al.\ for a model with a variable coupling for one link per
plaquette\cite{Berge_DGL}, and summarize our findings.

\newpage
\section{Background}

In this section we describe the model, discuss some quantities measured in
the MC simulations and their relation to the more convenient Coulomb gas
quantities, and shortly describe some analyses to be applied to the MC data.

\subsection{Model}

The model is defined through the partition function
\begin{displaymath}
  Z = \int_0^{2\pi} \prod_i \frac{d\theta_i}{2\pi} e^{-\beta H},
\end{displaymath}
where the Hamiltonian for a frustrated system is given by
\begin{displaymath}
  H = \sum_{\langle ij\rangle} U(\theta_i - \theta_j - A_{ij}).
\end{displaymath}
In the present case -- the Villain version of the \FFXY\ model -- the spin
interaction $U(\phi)$, is given by
\begin{displaymath}
  e^{-\beta U(\phi)} = \sum_{n=-\infty}^{\infty} e^{-\beta J(\phi-2\pi n)^2/2}
\end{displaymath}
where $\phi$ is an angular difference between nearest neighbors.  In the
Hamiltonian above $A_{ij}$ is the vector potential, and the frustration is
given by the rotation of $A_{ij}$:
\begin{displaymath}
  f_\r = \frac{1}{2\pi}\D \times \A_\r \equiv \frac{1}{2\pi}
  \left(A_{\r+\xhat/2}^y - A_{\r-\xhat/2}^y - A_{\r+\yhat/2}^x
    + A_{\r-\yhat/2}^x\right).
\end{displaymath}
Full frustration, $f = 1/2$, may e.g.\ be obtained by setting $A_\r^y = 0$
everywhere and $A_\r^x = \pi$ at every second row and zero otherwise. Here
we introduce the discrete difference operator, $\D = (D_x, D_y)$, $D_\mu
f_\r = f_{\r+\muhat/2} - f_{\r-\muhat/2}$, and $A^x$ and $A^y$ for the
vector potential at links in the $x$ and $y$ directions, respectively.
Associated with the discrete difference is $\tilde{k}$ which is obtained
from
\begin{displaymath}
  i\tilde{k}_x e^{i\kr} = D_x e^{i\kr} \;\;\;\Rightarrow \;\;\; \tilde{k}_x
  = 2\sin\frac{k_x}{2},
\end{displaymath}
and also gives $\tk^2 = 4 - 2\cos k_x - 2\cos k_y$.

\subsection{Measured quantities}

We now describe some of the quantities which are measured in the
simulations, and of central importance for the analyses in Secs.\
\Ref{Sec:Finite} and \Ref{Sec:Corr}.

\subsubsection{Helicity modulus}

The helicity modulus $\Upsilon$, is a measure of the quasi-long range
order in $XY$ models. It is defined from the increase in free energy due to
a small twist $\Delta$ across the system in one direction,
\begin{displaymath}
  \Upsilon = \parti{^2 F}{\Delta^2}.
\end{displaymath}
Written in this way, and with current $= \partial F/\partial\Delta$, the
helicity modulus may be interpreted as the proportionality constant between the
applied twist and the obtained macroscopic current.

In MC simulations the helicity modulus is obtained from the correlation
function\cite{Ohta_Jasnow}
\begin{displaymath}
  \Upsilon = J_0 - \frac{\beta}{L^2}\left<\left(\sum_\r U'(\phi_\r^x)
  \right)^2 \right>,
\end{displaymath}
where $J_0 = \left<U''(\phi) \right>$, and the sum in the second term is
over all links in one direction, here the $x$ direction.

\subsubsection{Vorticity}

Beside the helicity modulus the main quantity measured in our simulations is
the Fourier transform of the vorticity. The vorticity is defined in terms of
the rotation of the current $U'(\phi_{ij}) \equiv U'( \theta_i - \theta_j -
A_{ij})$ around a plaquette \cite{Miyashita_Shiba},
\begin{equation}
  \label{def.v}
  v = \frac{1}{2\pi J}(U'(\phi_{12}) + U'(\phi_{23}) +  U'(\phi_{34})
  + U'(\phi_{41})).
\end{equation}
The factor $2\pi$ in the denominator is chosen to give the zero-temperature
limit $v=\pm 1/2$. This follows from the angular difference $\phi=\pm
\pi/4$ in the ground state.  The steps in the simulations consist of
measuring $v_\r$ at each plaquette, Fourier transforming,
\begin{displaymath}
  v_\k = \sum_\r v_\r e^{-i\kr},
\end{displaymath}
and accumulating the Fourier components squared, $|v_\k|^2$.

It is also common to define the vorticity in terms of the angular
differences.  That corresponds to the chirality, in the context of \FFXY\ 
models.  However, an appealing feature of the vorticity defined in Eq.\ 
(\Ref{def.v}) is that it is related to some derivatives of the free energy.
This is also the reason for the existence of some {\em exact\/} relations
between the measured vortex correlations and the correlations in the 2D CG
with half-integer charges.  The chirality, on the other hand, is somewhat
peculiar in that it jumps discontinuously as a function of the angular
differences.

\subsubsection{The staggered magnetization}

For the study of the $Z_2$ degrees of freedom associated with the
symmetry of the anti-ferromagnetic Ising model, a convenient quantity
is the staggered magnetization
\begin{equation}
  M = \frac{2}{L^2} \left<\left| \sum_\r (-1)^{r_x + r_y} v_\r \right|\right>, 
  \label{def.Mv}
\end{equation}
where the sum is over all the plaquettes of the system, and the alternating
sign is include to take care of the checkerboard pattern. The factor of
$2/L^2$ is chosen to give $M = 1$ in a well-ordered system.  In an infinite
system this quantity has a finite value in the low-temperature phase and
goes to zero as $T_c$ is approached from below, but this sharp behavior is
considerably smoothed in the finite systems of the MC simulations.  Since
$(-1)^{r_x + r_y} = e^{-i(\pi,\pi)\cdot\r}$, $M$ is directly related to the
$\k = (\pi,\pi)$ component of the vorticity.  Also useful are some powers
of the staggered magnetization,
\begin{displaymath}
  M^p = \left(\frac{2}{L^2} \sum_\r (-1)^{r_x + r_y} v_\r \right)^p.
\end{displaymath}
Binder's cumulant in Sec.\ \Ref{Sec:Bg.Binder} is defined from
$\left<M^2\right>$ and $\left<M^4\right>$.

\subsection{Duality relation and the correlation function}
\label{Sec:Bg.dual-corr}

It has been argued that both the vortex interaction and the average
vorticity at a plaquette are temperature-dependent in the \FFXY\ model with
cosines interaction\cite{Olsson:xyff}.  To avoid this kind of complicating
factors one would rather have results from the CG with half-integer
charges, since both the average vorticity and the vortex interaction in
that model are manifestly temperature-independent.  However, since
simulations of that model are considerably more time-consuming, we instead
perform simulations of the spin model with Villain interaction, and make
use of an \emph{exact} relation between the measured vorticity correlations
and the corresponding correlations for the CG half-integer charges.  In
this section we shortly discuss the duality transformation, define the
$Z_2$ correlation function, $g(\r)$ and $g(\k)$, and derive the link
between our measured vorticity correlations and $g(\k)$.

In the Appendix we discuss the duality
transformation\cite{Jose_KKN,Vallat_Beck} applied to the \FFXY\ model.
This gives the Hamiltonian
\begin{equation}
  H^{\rm CG} = -4\pi^2 J \frac{1}{2}\sum_{\r,\r'} m_\r G(\r - \r') m_{\r'},
\end{equation}
where $m_\r$ are half-integer charges, $m_\r = \pm 1/2$, $\pm 3/2$, \ldots,
and $G(r)$ -- the lattices Green's function -- is the solution to $\D \cdot
G(\r) = \delta_\r$, with $G(0) = 0$ and proper boundary conditions. An
excellent approximation for $r\geq 1$ is $2\pi G(r) = \ln r + {\rm const}$.

It is now convenient to define the $Z_2$ correlation function in terms of
the CG charges $m_\r$.  For the correlation function in ordinary space we
write,
\begin{equation}
  \label{def.g}
  g(\r) = 4 (-1)^{r_x+r_y} \left<m_0 m_\r \right>,
\end{equation}
where the prefactors, again, are for normalization and the checkerboard
pattern in the well-ordered ground state.  In the low-temperature phase
this quantity has a finite value in the $r\rightarrow\infty$ limit, whereas
it approaches zero above $T_c$. The approaches to these limits are
exponential, governed by the correlation length $\xi$.  As discussed below,
the correlation length is, however, better determined from the Fourier
components $g(\k)$.

The Fourier expansion of the correlation function is
\begin{displaymath}
  g(\r) = \frac{1}{L^2} \sum_\k g(\k) e^{i\kr}
  = (-1)^{r_x+r_y} \frac{1}{L^2} \sum_\q g(\q) e^{i\q\cdot\r},
\end{displaymath}
where we introduce $\q = (\pi,\pi) - \k$ and $g(\q)$. Together with
Eq.\ (\Ref{def.g}) this gives
\begin{equation}
  g(\k) = \frac{4}{L^2}\left<m_\k m_{-\k} \right>.
  \label{gk.mk}
\end{equation}

A link between the vorticity correlation function and the corresponding
correlation function of CG charges may be obtained by considering two
different expressions for the wave-vector dependent helicity modulus
$\Upsilon(\k)$, which is equivalent to the dielectric function
$J/\epsilon(\k)$\cite{Minnhagen:review}.  In the Coulomb gas picture, with
the interaction $4\pi^2 J G(r) \sim 2\pi J\ln r$, $\Upsilon(\k)$ becomes
\begin{equation}
  \label{def.Upsk.m}
  \Upsilon(\k) = J - \frac{4\pi^2 J^2}{T L^2 \tk^2}
  \left<m_\k m_{-\k} \right>,
\end{equation}
whereas the same quantity in the $XY$ variables is
\begin{equation}
  \label{def.Upsk.v}
  \Upsilon(\k) = J_0 - \frac{4\pi^2 J^2}{T L^2 \tk^2}
  \left<v_\k v_{-\k} \right>.
\end{equation}
In the Appendix we show that these two expressions are the same derivative
of the free energy, which means that they have to be equal. This gives
the desired link between the correlation functions for our measured
vorticity and the half-integer variables of the CG model,
\begin{equation}
  \left<m_\k m_{-\k} \right> = \left<v_\k v_{-\k} \right> +
    \frac{T (J - J_0)}{J^2} \frac{L^2 \tk^2}{4\pi^2}.
  \label{m.v.rel}
\end{equation}
Together with Eq.\ (\Ref{gk.mk}) this gives the desired expression for the
correlation function $g(\k)$ in terms of the measured correlations
$\left<v_\k v_{-\k} \right>$.  This is the procedure used to determine
$g(\k)$ which is analyzed in Sec.\ \Ref{Sec:Corr}.

The last term in Eq.\ (\Ref{m.v.rel}) is under certain conditions very
small beside the vorticity correlation term $\left<v_\k v_{-\k} \right>$.
This is especially the case at $\q \approx 0$ for temperatures around
$T_c$, which means that the determinations of $\xi$ and $\nu$ in Sec.\ 
\Ref{Sec:Corr} would be influenced only very slightly by neglecting this
correction.

The relation between the CG correlations and the vorticity correlations in
ordinary space has been discussed in Ref.\ \onlinecite{Vallat_Beck}.  For
the case with Villain spin interaction the result was $\left<m_0 m_\r
\right> = \left<v_0 v_\r \right>$.  However, as seen in their derivation
this holds for $|\r| > 1$, only.  For general $\r$ the Fourier transform of
Eq.\ (\Ref{m.v.rel}) gives
\begin{equation}
  \left<m_0 m_\r \right> - \left<v_0 v_\r \right> = \left\{
  \begin{array}{ll}
    4 T(J- J_0)/ (2\pi J)^2, & \mbox{if $\r=0$}, \\
    - T(J- J_0)/ (2\pi J)^2, & \mbox{if $|\r| = 1$}, \\
    0, & \mbox{otherwise}.
  \end{array} \right.
  \label{m00.v00}
\end{equation}
In Sec.\ \Ref{Sec:Finite.check} the above relation for $\r = 0$ is used in
a consistency check.

\subsection{Analysis of $\Upsilon$}
\label{Sec:Bg.Ups}

In this section we discuss the size-dependence of the helicity modulus
$\Upsilon$ and its relation to the universal jump and Kosterlitz'
renormalization group (RG) equations.  We will focus on the dimensionless
quantity $T/\Upsilon$.

For the finite-size scaling analysis of $T/\Upsilon$ one assumes that the
size-dependence in this quantity is related to the behavior of a set of RG
trajectories\cite{Kosterlitz:74}.  The starting point in parameter space,
and thereby the relevant trajectory, is determined by the temperature.
These trajectories behave differently in the low- and high-temperature
phases.  In the low-temperature phase they terminate at finite values of
$T/\Upsilon$ whereas they continue to infinity, corresponding to $\Upsilon
\rightarrow 0$, in the high-temperature phase.

The last trajectory in the low-temperature phase ends at the universal
value $\TCG/\Upsilon \equiv T/(2\pi\Upsilon) = 1/4$. This means that the
helicity modulus for that very temperature in an infinite system is
$\Upsilon = 2T/\pi$. The jump of this quantity to zero is the well-known
universal jump\cite{Nelson_Kosterlitz,Minnhagen_Warren}.

The abrupt universal jump of an infinite system is, of course, not seen in
finite systems. Since $\Upsilon$ decreases with increasing system size, the
universal jump conditions $\Upsilon = 2T/\pi$ may, however, be used to
establish an upper limit for $\TKT$. The temperature obtained in that way
is a rigorous upper limit, since the universal jump condition constitutes
an absolute stability criterion.

The approach to the universal value, $\Upsilon\pi/(2T) = 1$, with
increasing system size, may also be used to examine the critical
properties.  From Kosterlitz' RG equations\cite{Kosterlitz:74} the
finite-size scaling relation for $\Upsilon_L$
becomes\cite{Weber_Minnhagen:88}
\begin{equation}
  \label{weber-scaling}
  \frac{\Upsilon_L\pi}{2T} = 1 + \frac{1}{2(\ln L + l_0)}.
\end{equation}
Kosterlitz' RG equations are expected to be valid only in the limit of low
vortex density. This means that the above finite size scaling relation is
expected to be valid only at low renormalized vortex density. Accordingly,
$\Upsilon_L\pi/(2T)$ should be not too far from unity -- $\Upsilon$
renormalized out to length scale $L$ should not be too far from the
fully renormalized $\Upsilon$ out to infinity.  This implies the dilute
limit for sizes bigger than $L$.

The same idea may also be used both above and below $\TKT$. A more complete
discussion is given in Ref.\ \cite{Olsson:Kost-fit}.  Close to $\TKT$ we
expect \cite{Minnhagen:review,Olsson:Kost-fit}
\begin{eqnarray}
  \frac{\Upsilon_L\pi}{2T} & = & 1 + c\coth[2c(\ell_0 + \ln L)], \;\;\; T <
  T_c, \label{Kost.lo} \\ \frac{\Upsilon_L\pi}{2T} & = & 1 - c\cot
  [2c(\ell_0 - \ln L)], \;\;\; T > T_c. \label{Kost.hi},
\end{eqnarray}
where $\ell_0$ and $c$ are free parameters to be determined from the fits.
Eq.\ (\Ref{weber-scaling}), is the $c \rightarrow 0$ limit of Eq.\ 
(\Ref{Kost.lo}).  $c$ vanishes as $\TKT$ is approached from below or
above as\cite{Kosterlitz:74}
\begin{displaymath}
  c = B \sqrt{\left|T/\TKT - 1 \right|}.
\end{displaymath}
In the high-temperature phase $\ell_0$ is identified with the logarithm of
the screening length, $\lambda$.  In the immediate vicinity of $\TKT$ the
temperature dependence of $\ell_0$ should therefore be given by Kosterlitz'
expression\cite{Kosterlitz:74}
\begin{equation}
  \ell_0 = \frac{C}{\sqrt{T/\TKT - 1}} + {\rm const},
  \label{l0.sqrt}
\end{equation}
where\cite{Minnhagen:review}
\begin{equation}
  \label{C.B}
  C = \frac{\pi}{2 B}.
\end{equation}

\subsection{Binder's cumulant}
\label{Sec:Bg.Binder}

Binders' cumulant is a convenient quantity that, in most cases, facilitates
determinations of both the critical temperature and the correlation length
exponent $\nu$.  Even though the quantity was originally presented in terms
of averages over blocks of different size in a single simulation with a
fixed total system size\cite{Binder:cumulant}, it may also be used with
data from systems of different size. Binder's cumulant is obtained from
some moments of the order parameter,
\begin{equation}
  U = 1 - \frac{\left<M^4\right>}{3\left<M^2 \right>^2}.
  \label{UL}
\end{equation}
The crucial property of $U$ is its size-independence precisely at $T_c$.
Therefore, plotting $U$ versus temperature for several different sizes is
expected to give an unique crossing point at the critical temperature.
Furthermore, the correlation length exponent $\nu$ may be determined by
plotting the data against $(T-T_c) L^{1/\nu}$.  The correct value of $\nu$
is expected to give a collapse of that data onto a single curve.  In
practice there are, however, often corrections to scaling which make the
conclusions from this kind of analysis less direct and precise.

\subsection{Boundary conditions}\label{Sec:Bg.bc}

It has recently been pointed out that periodic boundary conditions (PBC's)
in the $XY$ model may be generalized by including twist fluctuations along
the $x$ and $y$ directions in the system -- fluctuating boundary conditions
(FBC's) \cite{Olsson:self-cons,Olsson:self-cons.long}. There are several
advantages with considering such a generalization. First, it is with these
boundary conditions that the Villain version of the $XY$ model is exactly
dual to the CG with periodic boundary conditions. Second, the finite size
effects in several quantities work in the opposite way after the inclusion
of these twists.  Finally, with a self-consistently chosen amplitude of
these twists, the finite size effect on the correlation function turns out
to be virtually eliminated.

The self-consistent boundary conditions do not seem to be applicable in the
fully frustrated case. This is possibly an effect of the $Z_2$
fluctuations.  It is, however, possible to obtain the correlation function as in
an infinite system by taking the average of data for PBC's and FBC's, cf.\ 
Fig.\ 1 in Ref.\ \cite{Olsson:self-cons}.  This technique works up to, and
possibly slightly above $\TKT$.  At higher temperatures both sets of data
go down with increasing lattice size, which makes it considerably more
difficult to extract any result for the thermodynamic limit.

\newpage
\section{Finite size analyses}
\label{Sec:Finite}

The different methods to analyze MC data may, generally speaking, be
divided into two classes. The most obvious one is to calculate the
correlation functions and determine the correlation length and the
associated exponents from this kind of data.  In this kind of analyses
one is interested in the behavior of an infinite system and, accordingly,
the finite-size effects are undesired complications.  This kind of analyses
is the subject of Sec.\ \Ref{Sec:Corr}.

The second class of methods instead \emph{take advantage} of the finite-size
dependence in the MC data.  This is generally a more efficient approach to
analyzing the critical behavior.  In this Section we employ some techniques
that make use of the finite-size dependence in various ways.

After a short description of the simulations and some checks employed to
validate the results, we focus on results from the universal jump condition
in Sec.\ \Ref{Sec:Finite.univ}.  In Sec.\ \Ref{Sec:Finite.TKT} we perform
finite-size scaling analyses of the helicity modulus $\Upsilon$ both right
at $\TKT$ and in the immediate neighborhood around $\TKT$. With this
determined value for $\TKT$ we then take a closer look at the data from the
universal jump condition in Sec.\ \Ref{Sec:Finite.TKTL}.

To obtain a reference temperature we then apply finite-size scaling
analysis of Binder's cumulant at $T_c$.  Just as in the related models this
kind of analysis gives $\nu < 1$.  As suggested in Ref.\ \cite{Olsson:xyff}
this seems to be an artifact of the presence of a finite screening length
$\lambda$ associated with the nearby KT transition.

\subsection{Monte Carlo simulations}

The Monte Carlo simulations were performed with the ordinary Metropolis
algorithm with sequantial sweeps over the lattices.  One such sweep with
one trial update per spin is called a MC step.  For most of the data there
were four MC steps between consecutive measurements. But since it was noted
that a major part of the computer time, especially on the large lattices,
was used in the Fast Fourier Transform of the measured vorticity $v_\r$,
the simulations for $L=128$ and 256 for determinations of $\xi$ were
performed with as much as 64 MC steps between consecutive measurements.
For the latter data the number of MC steps are given in Table
\Ref{tab:Corr}.

In Sec.\ \Ref{Sec:Corr} we make use of MC data from the 2D Ising model as a
convenient testing ground for the methods used to analyze the \FFXY\ model.
These simulations are performed with Wolff's cluster
algorithm\cite{Wolff:89a}.  All the simulations were done on a set of
DEC-alpha workstations.

\subsection{Monte Carlo data}
\label{Sec:Finite.check}

A MC study is, of course, never more reliable than the underlying data.  It
is therefore essential to check that the program, indeed, does provide
correct data.  This may be done either by comparing with previously
published results or by making use of some consistency tests.

To the best of our knowledge there is no published MC data to compare with
for the Villain version of the fully frustrated $XY$ model with ordinary
PBC's. For the case with FBC's it is, however, possible to compare with data
for the half-integer CG\cite{J-R.Lee:94}. For $L=8$ and $T/J = 0.82$ our
simulations give Binder's cumulant, $U = 0.5786$. This temperature
corresponds to $\TCG = 0.1305$, and as expected our value for $U$ lies
right in-between the values for $U$ at $\TCG = 0.130$ and 0.131 in Fig.\ 3
of Ref.\ \cite{J-R.Lee:94}. As a second test we compare the values of
$\Upsilon(k=2\pi/L)$ for $L = 16$ and $T/J = 0.82$. Again, the value
obtained from our simulations, 0.556788, is in good agreement with the
corresponding values in Fig.\ 4 of Ref.\ \cite{J-R.Lee:94}.

For the bulk of our data, obtained with ordinary PBC's we have to resort to
internal consistency tests. One such test is suggested by the analogy with
the CG with half-integer charges. In that case, the charges $m = \pm 1/2$
give $\left<m^2 \right> = 1/4$. Actually, the value $1/4$ turns out to be a
lower bound since the CG also includes charges of non-lowest order, i.e.\ 
$m = \pm 3/2$. For our measured quantity $v^2$, there is no such simple
result, but as discussed above there is an exact relation between these two
quantities, Eq.\ (\Ref{m00.v00}).

\begin{figure}
  \hspace{2cm}\input{gg}\put(0,0){%
  \epsfig{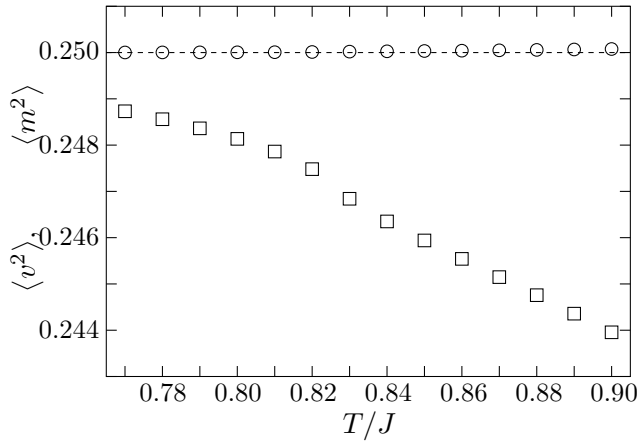}}\end{picture}
  \caption{\small The average vorticity squared for a single
  plaquette $\left<v^2 \right>$, together with the corresponding
  Coulomb-gas quantity $\left<m^2 \right>$ as functions of temperature.
  That the latter quantity, obtained through Eq.\ (\Ref{m00.v00}), is close
  to $1/4$ is a consistency test of the simulations.  The small deviations
  from $1/4$ give information about the density of non-lowest charges,
  $m=\pm 3/2$, in the corresponding Coulomb gas.}\label{fig-gg}
\end{figure}

The behavior of both $\left<v^2 \right>$ (squares) and $\left<m^2 \right>$
from Eq.\ (\Ref{m00.v00}) (circles) is shown in Fig.\ \Ref{fig-gg}.
Whereas $\left<v^2 \right>$ is seen to decrease with temperature we find
that $\left<m^2 \right>$ indeed is very close to $1/4$.  More precisely,
the results are $\left<m^2 \right> = 0.250000$ at $T/J = 0.4$, 0.250003 at
$T/J = 0.77$, and 0.250078 at $T/J = 0.9$.  This constitutes a confirmation
of the correctness of the MC data.  From the minute deviations of
$\left<m^2 \right>$ from $1/4$ it is also possible to obtain estimates of
the fraction of plaquettes with non-lowest order charges in an equivalent
CG\cite{Wallin-help}.  For the temperature interval in Fig.\ \Ref{fig-gg}
the data indicates that this fraction would be from $1.5\times 10^{-6}$ to
$39 \times 10^{-6}$.

\subsection{Universal jump condition}\label{Sec:Finite.univ}

In this section we make use of the universal jump condition to obtain both
an upper limit of $\TKT$ and a strong evidence in favor of two distinct
transitions.  Since the universal jump condition is an absolute stability
requirement, we believe the argument of the present section to be
especially free of objections.  Whereas the more precise results in the
later sections are obtained on the basis of additional assumptions, the
direct use of the universal jump condition is particularly clean.

For the following discussion we introduce the size-dependent transition
temperature $\TKTL$, as the temperature where the helicity modulus for system
size $L$ is equal to the universal value, $\Upsilon_L = 2\TKTL/\pi$.

\subsubsection{Upper limit of $\TKT$}
\label{Sec:Finite.univ.limit}

The universal jump condition was applied to the fully frustrated $XY$ model
with cosines interaction in Ref.\ \cite{Lee_Lee} to establish an upper
limit for the KT temperature. From the intersection of the MC data for $L =
128$ with the universal value, these authors found, as discussed in the
Introduction, $\TKT^{(128)}/J = 0.449(1)J$, clearly below the values of $T_c$
determined with finite size scaling.

\begin{figure}
  \hspace{2cm}\input{Ups.t}\put(0,0){%
  \epsfig{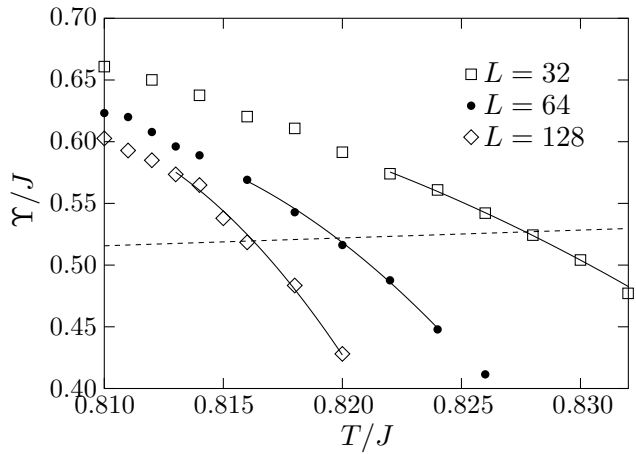}}\end{picture}
  \caption{\small Determinations of the size-dependent KT
  temperatures $\TKTL$. The dashed line is the universal jump condition.
  The solid lines are second order polynomials in $T/J$ obtained from fits
  to the MC data.}\label{fig-Ups.t}
\end{figure}

\begin{figure}
  \hspace{2cm}\input{TKT.L}\put(0,0){%
  \epsfig{file=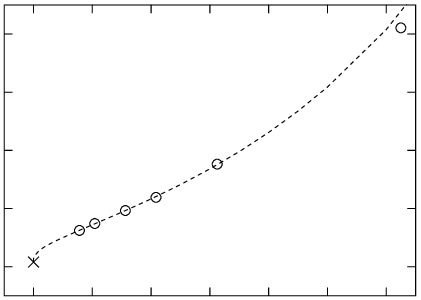}}\end{picture}
  \caption{\small The size-dependence of $\TKTL$.  The cross is
  $\TKT$ from Sec.\ \Ref{Sec:Finite.TKT} and the dashed line is from the
  analysis in Sec.\ \Ref{Sec:Finite.TKTL}.}\label{fig-TKT.L}
\end{figure}

The same approach with the data for the Villain version is shown in Fig.\ 
\Ref{fig-Ups.t}. The upper limit obtained for $L=128$ is $\TKT^{(128)}
\approx 0.816 J$.  Figure \Ref{fig-TKT.L} shows the size-dependence of
$\TKTL$. The dashed line is from the analysis in Sec.\ 
\Ref{Sec:Finite.TKTL}.

\subsubsection{Two transitions}
\label{Sec:Finite.univ.two}

We now turn to the argument for two distinct transitions based solely on the
universal jump condition\cite{Nelson_Kosterlitz,Minnhagen_Warren}.

The starting point is that the staggered magnetization for an infinite
system, $M_\infty$, vanishes at the $Z_2$ transition temperature $T_c$. To
establish the existence of two distinct transitions it is therefore
sufficient to examine $M_\infty$ right at $\TKT$.  A non-zero value of
$M_\infty(\TKT)$ would be an unequivocal demonstration of Ising order,
which then implies that this order is lost at a higher temperature, $T_c >
\TKT$. The determination of $M_\infty(\TKT)$ at first seems very difficult
since, beside the usual problem of approaching the thermodynamic limit, the
value of $\TKT$ has to be known with high precision. The universal jump
criterion used so far, is only capable of yielding upper limits.

A way around both these difficulties at the same time is to focus on
$M_L(\TKTL)$, the staggered magnetization at finite lattices at the
size-dependent KT temperatures, and examine the behavior of this quantity
as a function of system size, $L$.  The point is that the desired quantity
$M_\infty(\TKT)$ is the large-$L$ limit of $M_L(\TKTL)$, and that the
staggered magnetization is readily determined for each $L$.

\begin{figure}
  \hspace{2cm}\input{m.L}\put(0,0){%
  \epsfig{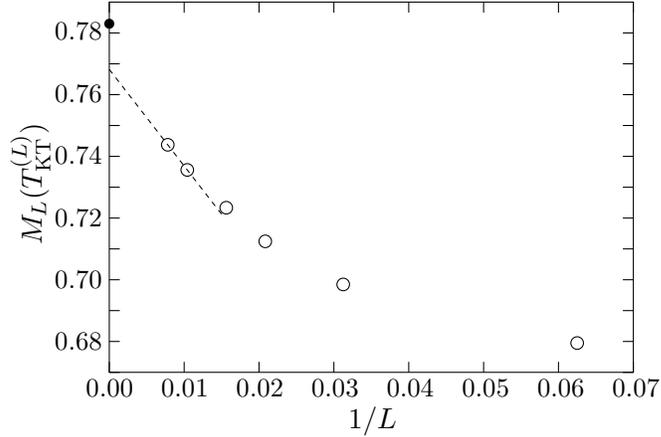}}\end{picture}
  \caption{\small Evidence for two distinct transitions. The plotted
  quantity is known to approach $M_\infty(\TKT)$ in the limit
  $L\rightarrow\infty$.  The dashed line shows a naive extrapolation,
  whereas the dot is this quantity from Sec.\ \Ref{Sec:Corr.bc} obtained
  with the value of $\TKT$ from Sec.\ \Ref{Sec:Finite.TKT}.}\label{fig-m.L}
\end{figure}

The results from this analysis is shown in Fig.\ \Ref{fig-m.L}.  The figure
shows that the staggered magnetization at $\TKTL$ is an \emph{increasing}
function of lattice size.  The figure gives 0.744 as a lower limit of
$M_\infty(\TKT)$, and a naive extrapolation, that neglects the curvature,
suggests $M_\infty(\TKT) > 0.768$.

We consider this to be a very strong argument that the $Z_2$ order persists
at the KT transition temperature.  For the opposite to be true, this
increasing trend toward a finite value of $M_\infty(\TKT)$ should change to
a decreasing trend down to zero. Even though this possibility could never
be ruled out from the data for finite systems alone, such a change in trend
seems very unlikely.  Furthermore, the more detailed analysis in the
following sections yields $M_\infty(\TKT) = 0.783(2)$, entirely consistent
with the increasing trend in Fig.\ \Ref{fig-m.L}.  It should also be noted
that this line of evidence does not depend on the assumption of an
\emph{universal} jump.  The argument holds equally well with a jump
$\Upsilon\pi/(2T) = g$, $g> 1$.  ($g<1$ is excluded by stability.)

\begin{figure}
  \hspace{2cm}\input{m.q}\put(0,0){%
  \epsfig{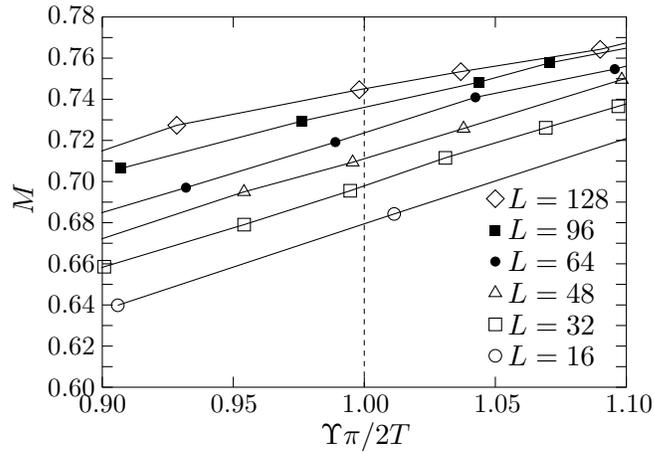}}\end{picture}
  \caption{\small The determination of $M_L(\TKTL)$.  Since $\TKTL$
  is the temperature for which $\Upsilon_L\pi/(2T) = 1$, the desired
  quantity is obtained from the crossing of $M$ for the different sizes
  with the dashed line.}\label{fig-m.q}
\end{figure}

The actual determination of $M_L(\TKTL)$ is illustrated in Fig.\ 
\Ref{fig-m.q}.  $M_L(\TKTL)$ is obtained directly from the value of $M_L$
where $\Upsilon_L = 2T/\pi$ (dashed line in Fig.\ \Ref{fig-m.q}).

\subsection{Kosterlitz-Thouless transition}
\label{Sec:Finite.TKT}

The purpose with this section is first to determine the Kosterlitz-Thouless
temperature, $\TKT$, and, second, to examine the behavior closely below and
above this temperature. The method employed is finite size scaling
analysis of the helicity modulus as discussed in Sec.\ \Ref{Sec:Bg.Ups}.
The basic idea is that the size-dependence of the helicity modulus
$\Upsilon$, at and in the vicinity of $\TKT$, may be obtained from Kosterlitz
RG equations\cite{Kosterlitz:74} as given by Eqs.\ (\Ref{weber-scaling}),
(\Ref{Kost.lo}) and (\Ref{Kost.hi}).

The analysis of the helicity modulus $\Upsilon$, is based on a large amount
of MC data. In order to make efficient use of the data and get $\Upsilon_L$
as continuous functions of $T$, we first determine the helicity modulus as
second order polynomials, one for each $L$, in $\tau = T/J - 0.8107$:
\begin{equation}
  \frac{\Upsilon_L\pi}{2 T} = \alpha_L + \beta_L \tau + \gamma_L
  \tau^2.
  \label{Ups.smooth}
\end{equation}
These second order expansions are only expected to be valid within rather
narrow temperature intervals. We therefore only include data in the fits
for temperatures $|\tau| < T_{\rm range}/J$, where $T_{\rm range}$
decreases with increasing $L$.  The parameters from this analysis together
with the size of the temperature intervals are shown in Table
\Ref{tab:Ups.tau}.
\begin{table}[t] \begin{center}
  \label{tab:Ups.tau}
  \begin{tabular}[l]{rrrrr}
    $L$ & $\alpha_L$ & $\beta_L$ & $\gamma_L$ & $T_{\rm range}/J$ \\ 
    \hline
     16 & 1.3932(17) & -8.545 & -19.9  & 0.021 \\
     24 & 1.3179(20) & -11.06 & -143.6 & 0.015 \\
     32 & 1.2731(12) & -12.86 & -207.4 & 0.01 \\
     48 & 1.2235(13) & -15.75 & -190.8 & 0.008 \\
     64 & 1.1989(13) & -17.55 & -352.1 & 0.0065 \\
     96 & 1.1714(15) & -20.20 & -439.0 & 0.0051 \\
    128 & 1.1578(20) & -21.30 & -304.1 & 0.0045 \\
    \hline
    \end{tabular}
  \end{center}
  \caption{Parameters from fitting MC data for the helicity modulus to 
    Eq.\ (\protect\ref{Ups.smooth}).  The data included in the fits are
    restricted to $|T/J - 0.8107| < T_{\rm range}/J$.}
\end{table}

\subsubsection{Determination of $\TKT$}

\begin{figure}
  \hspace{2cm}\input{Ups.Lmin}\put(0,0){%
  \epsfig{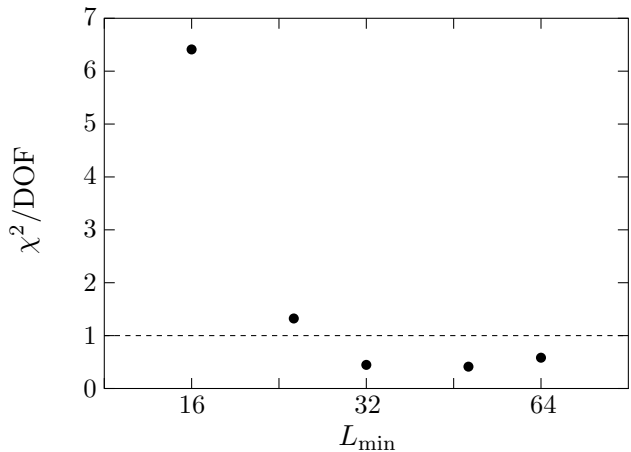}}\end{picture}
  \caption{\small The quality of the fit from fitting
  $\Upsilon_L$ to Eq.\ (\Ref{weber-scaling}).  $\Lmin$ is the smallest size
  included in the fit.  For a good fit one expects $\chi^2$/DOF $\approx
  1$.  The figure indicates that the fit is not very sucessful when
  including small lattices, but becomes acceptable for $\Lmin =
  32$.}\label{fig-Ups.Lmin}
\end{figure}

We now apply the finite size scaling relations for the helicity modulus as
discussed in Sec.\ \Ref{Sec:Bg.Ups}. Since these relations in this system
only are expected to be valid for fairly large lattices we first follow the
procedure in Ref.\ \cite{Olsson:Kost-fit} and perform the analysis with
systems of size $L = \Lmin$ through 128 and various values for $L_{\rm
  min}$.  The errors in the fits are shown in Fig.\ \Ref{fig-Ups.Lmin}. On
the basis of this analysis we conclude that $\Lmin = 32$ does give a good
fit.  This is the same choice as for the \FFXY\ model with cosines
interaction in Ref.\ \cite{Olsson:xyff}.

\begin{figure}
  \hspace{2cm}\input{Ups.L}\put(0,0){%
  \epsfig{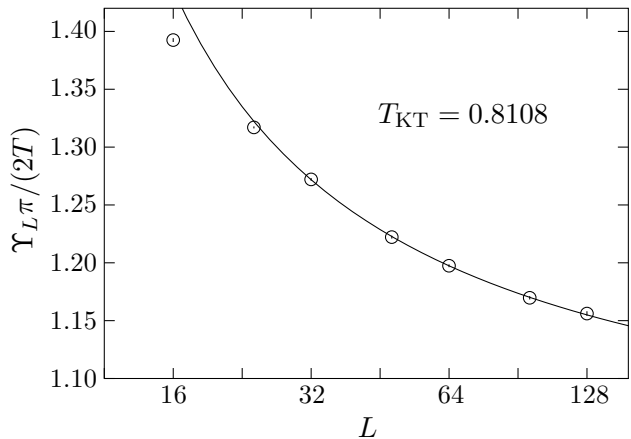}}\end{picture}
  \caption{\small $\Upsilon_L$ versus lattice size for $\TKT =
  0.8108 J$. The solid line is Eq.\ (\protect\Ref{weber-scaling}) with $l_0
  = -1.63$. The good fit is strong evidence for an ordinary KT
  transition.}\label{fig-Ups}
\end{figure}

Figure \Ref{fig-Ups} shows the good fit of the MC data to Eq.\ 
(\Ref{weber-scaling}) obtained by adjusting $\TKT$ and $l_0$.  The obtained
value for the KT temperature is $\TKT/J = 0.8108(1)$. We consider the good
fit to Eq.\ (\Ref{weber-scaling}) to be very strong evidence for an
ordinary KT transition.

Note that $\TKT/J = 0.8108$ is well below the upper limit $T/J = 0.816$
from the universal jump criterion in Fig.\ 3. It is also slightly lower
than what a simple linear extrapolation of the four lowest points to $1/L =
0$ in Fig.\ \Ref{fig-TKT.L} would suggest.

\subsubsection{Finite size scaling around $\TKT$}

We now shortly discuss the critical behavior in the immediate vicinity
of $\TKT$.  The approach closely follows Ref.\ \cite{Olsson:Kost-fit}.  

In fitting our MC data to Eqs.\ (\Ref{Kost.lo}) and (\Ref{Kost.hi}) we fix
the temperature, calculate $\Upsilon_L\pi/(2T)$ from the parameters in
Table \Ref{tab:Ups.tau} and adjust $c$ and $l_0$ to get the best possible
fit.  $l_0$ is the logarithm of the screening length, $\lambda$, in
the high-temperature phase. Below $\TKT$ this quantity has no such direct
interpretation. The results from this kind of fitting for several
temperatures around $\TKT$ are shown in Table \Ref{tab:c.l0}.

\begin{figure}
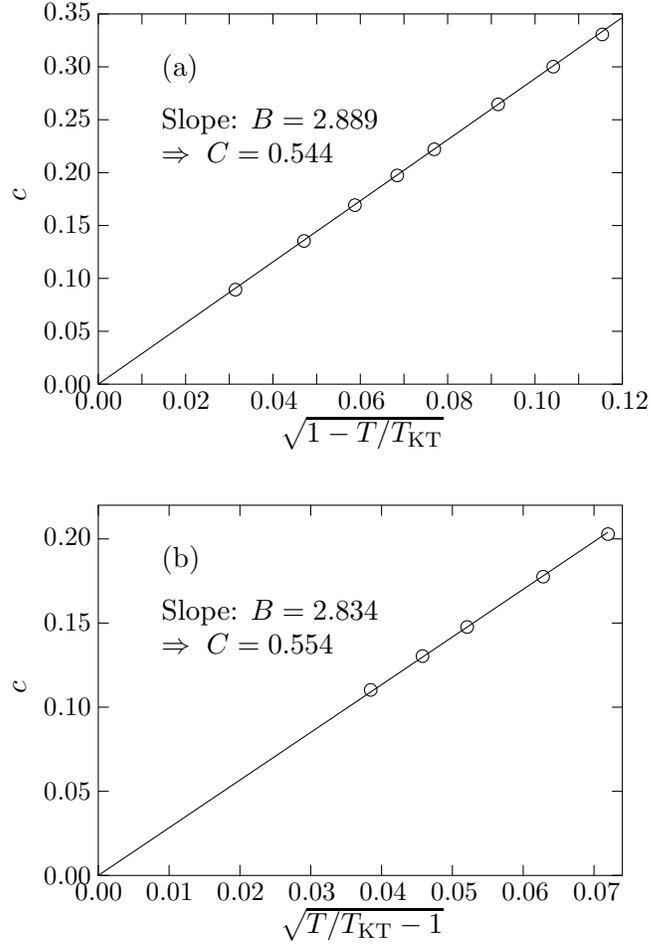

  \hspace{2cm}\input{lo.c.sqrt}\put(0,0){%
  \epsfig{file=lo.c.sqrt.eps}}\end{picture}
  \par\vspace{0.5cm}\noindent
  \hspace{2cm}\input{hi.c.sqrt}\put(0,0){%
  \epsfig{file=hi.c.sqrt.eps}}\end{picture}
  \caption{\small The temperature
  dependence of the parameter $c$ in Table \Ref{tab:c.l0}.  The data are
  obtained by fitting $\Upsilon_L$ to Eqs.\ (\Ref{Kost.lo}) and
  (\Ref{Kost.hi}), respectively.  (a) Low temperatures, $T <
  \TKT$.  In this case $c = \Upsilon_\infty\pi/(2T) - 1$, and the figure
  therefore illustrates the approach of $\Upsilon_\infty$ to the universal
  value as $\protect\sqrt{1-T/\TKT}$.  At high temperatures, panel (b), $c$
  has a similar square-root cusp, though it is no longer related to
  $\Upsilon_\infty$.}\label{fig-c.sqrt}
\end{figure}

\begin{figure}
  \hspace{2cm}\input{hi.l0.sqrt}\put(0,0){%
  \epsfig{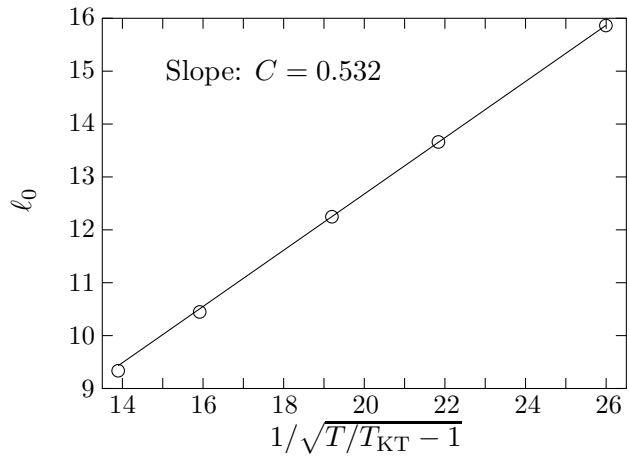}}\end{picture}
  \caption{\small The temperature dependence of the screening
  length above $\TKT$, predicted by Kosterlitz, Eq.\ (\Ref{l0.sqrt}).
  Beside an additive constant, $\ell_0$ is equal to the logarithm of the
  screening length.  This is a very direct determination of the slope $C$.
  The data, which is also listed in Table \Ref{tab:c.l0}, was obtained by
  fitting $\Upsilon_L$ to Eq.\ (\protect\Ref{Kost.hi}).}\label{fig-hi.l0.sqrt}
\end{figure}

Figures \Ref{fig-c.sqrt} show the temperature dependence of $c$.  Just as
in the analysis of the ordinary $XY$ model the values of the slopes $B$
obtained from the low- and high-temperature data are, within statistical
errors, the same.  Also shown in the figures are the corresponding values
of $C$ from Eq.\ (\Ref{C.B}).  The slope of $\ell_0$ versus $1/\sqrt{T/\TKT
  - 1}$ in Fig.\ \Ref{fig-hi.l0.sqrt}, is in good agreement with the values
from Figs.\ \Ref{fig-c.sqrt}, and our estimate for the slope becomes $C =
0.54\pm 0.02$. In Secs.\ \Ref{Sec:Finite.TKTL} and \Ref{Sec:Corr.lambda} we
will obtain different values for $C$, but we consider the present
determination to be the more reliable one for two reasons. First, it does
build on excellent agreements with predictions from the Kosterlitz' RG
equations and, second, in contrast to other determinations, this method
does probe the behavior in the immediate vicinity of $\TKT$.

\begin{table}[t]
  \begin{center}
    \leavevmode
    \begin{tabular}[l]{cccc}
      $T/J$ & $c$ & $\ell_0$ & $\chi^2/{\rm DOF}$ \\
      \hline
      0.8   &  0.3306 &  -1.5659 &  0.46 \\
      0.802 &  0.3002 &  -1.5904 &  0.46 \\
      0.804 &  0.2646 &  -1.6109 &  0.45 \\
      0.806 &  0.2221 &  -1.6254 &  0.38 \\
      0.807 &  0.1973 &  -1.6287 &  0.25 \\
      0.808 &  0.1693 &  -1.6293 &  0.10 \\
      0.809 &  0.1354 &  -1.6293 &  0.06 \\
      0.81  &  0.0894 &  -1.6286 &  0.10 \\
      \hline
      0.812 &  0.1103 &  15.86  &  0.68 \\
      0.8125&  0.1304 &  13.66  &  1.13 \\
      0.813 &  0.1476 &  12.25  &  1.66 \\
      0.814 &  0.1775 &  10.45  &  1.98 \\
      0.815 &  0.2029 &   9.34  &  1.60 \\
      \hline
    \end{tabular}
    \caption{Parameters from the fitting of MC data to Eqs.\
      (\protect\Ref{Kost.lo}) and (\protect\Ref{Kost.hi}) below and above
      $\TKT$, respectively.}
    \label{tab:c.l0}
  \end{center}
\end{table}

\subsection{Size-dependence of $\TKTL$}
\label{Sec:Finite.TKTL}

In Sec.\ \Ref{Sec:Finite.univ} we made use of the universal jump condition
to determine a kind of size-dependent KT temperatures $\TKTL$ as
upper bounds for $\TKT$.  From Fig.\ \Ref{fig-TKT.L}, it seemed difficult
to extrapolate such data to the thermodynamic limit.

\begin{figure}
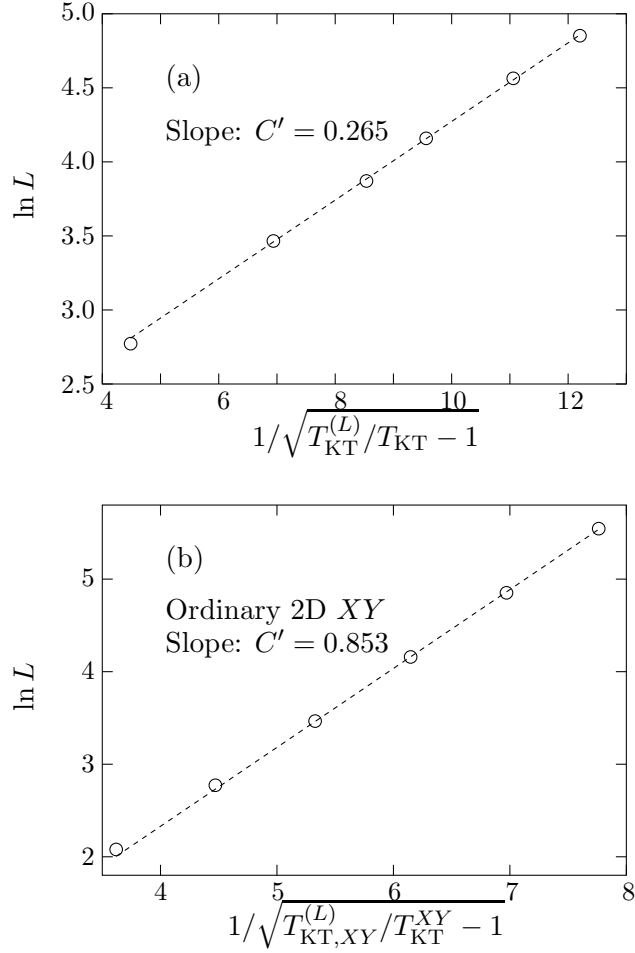

  \hspace{2cm}\input{TKT.L.line}\put(0,0){%
  \epsfig{file=TKT.L.line.eps}}\end{picture}
  \par\vspace{0.5cm}\noindent
  \hspace{2cm}\input{XY.TKT.L}\put(0,0){%
  \epsfig{file=XY.TKT.L.eps}}\end{picture}
  \caption{\small The size dependence
  of $\TKTL$. (a) For the \FFXY\ model, with lattice sizes $L = 16$, 32,
  48, 64, 96, and 128.  The value for the transition temperature is taken
  from the determination in Sec.\ \Ref{Sec:Finite.TKT}.  (b) The
  corresponding quantity in the ordinary $XY$ model with no frustration.
  This is with $\TKT^{XY}/J = 0.8921$\protect\cite{Olsson:Kost-fit}.  Note
  that the slope $C'$ in these cases are entirely different from the
  corresponding slopes from the temperature dependence of the screening
  length.}\label{fig-TKT.L.line}
\end{figure}

We now demonstrate that the size-dependence of $\TKTL$ has the same form as
the Kosterlitz' expression for the temperature dependence of the
correlation length, i.e.
\begin{equation}
  \ln L = {\rm const} + \frac{C'}{\sqrt{\TKTL/\TKT - 1}}.
  \label{lnL.sqrt}
\end{equation}
Figure \Ref{fig-TKT.L.line}(a) shows the approach of $\TKTL$ to $\TKT =
0.8108 J$ with increasing $L$. The points do, indeed, fall on a straight
line. This is the same dashed line as in Fig.\ \Ref{fig-TKT.L}. It should
be noted that it not seems possible to link this behavior directly to the
divergence of the screening length, $\lambda$.  In the present approach the
slope is $C' \approx 0.265$, whereas the same constant in the temperature
dependence of $l_0$ yielded $C \approx 0.54$ which is about twice as big.

This size-dependence of $\TKTL$ is also found in the ordinary 2D $XY$ model
with no frustration. This is illustrated in Fig.\ \Ref{fig-TKT.L.line}(b).
Here the slope is $\approx 0.85$, and, again, the temperature dependence of
the characteristic length gives a slope that is about twice as
big\cite{C-values}.

\subsection{Binder's cumulant}
\label{Sec:Finite.Binder}

As discussed in the Introduction the existence of two transitions close to
each other may give rise to problems with ordinary finite size scaling. The
evidence for two transitions given in the previous Section, strongly
implies that this actually is the case for the \FFXY\ model.  This means
that the critical properties not are accessible with finite size scaling at
$T_c$ unless the systems employed are considerably larger than the
correlation length associated with the other (here the KT) transition.

The purpose with the present finite size scaling analysis is therefore not
to extract the correct critical behavior, but rather to provide a reference
temperature and to verify that the \FFXY\ with Villain interaction indeed
does behave in a way that is similar to the more studied \FFXY\ model with
cosines interaction.

\begin{figure}
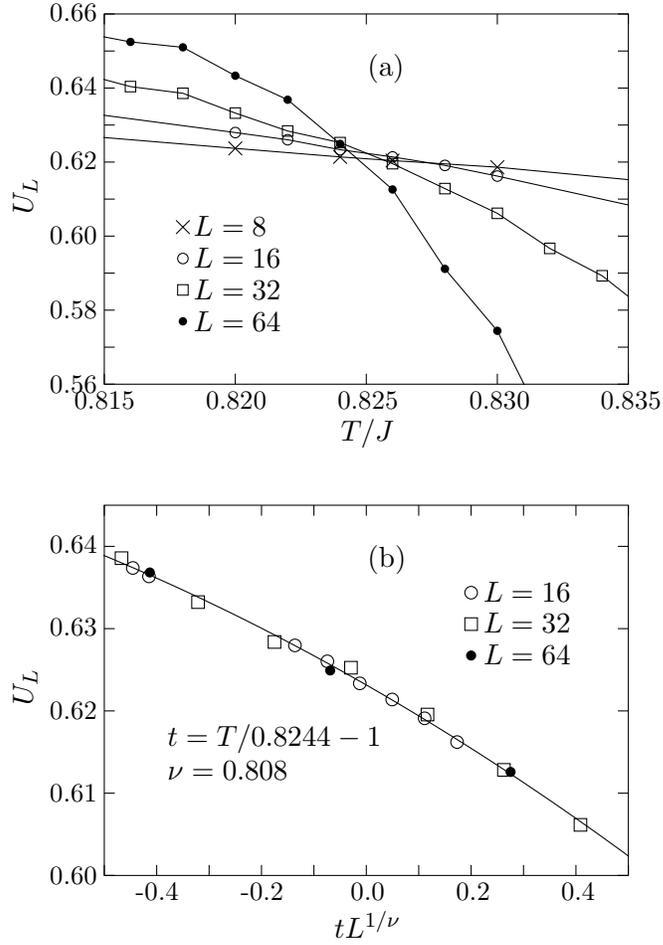

  \hspace{2cm}\input{binder}\put(0,0){%
  \epsfig{file=binder.eps}}\end{picture}
  \par\vspace{0.5cm}\noindent
  \hspace{2cm}\input{binder.fit}\put(0,0){%
  \epsfig{file=binder.fit.eps}}\end{picture}
  \caption{\small Binder's cumulant,
  $U_L$, for different lattice sizes.  (a) $U$ is more or less
  size-independent for $L \geq 16$ at $T/J \approx 0.824$.  (b) An
  attempted data collaps.  Just as in the \FFXY\ model with cosines
  interaction this kind of analysis suggests $\nu<1$.}\label{fig-binder}
\end{figure}

Figure \Ref{fig-binder}(a) shows Binder's cumulant versus temperature for $L =
8$, 16, 32, and 64.  As discussed in Sec.\ \Ref{Sec:Bg.Binder}, $U$ is
expected to be size-independent right at the critical temperature.  This is
not quite borne out by the data.  A close look reveals that the crossing
points move slowly to lower temperatures for larger system sizes.  For the
pairs of lattice sizes $L = 8$, 16, $L = 16$, 32, and $L = 32$, 64, the
crossing temperatures are 0.827, 0.825, and 0.824, respectively, though the
two last temperatures are within the statistical uncertainties.  This is in
good agreement with $2\pi\times 0.1315(3) = 0.826(2)$ obtained from
simulations of the CG with half-integer charges on lattices with size $L =
10$ -- 24 in Ref.\ \cite{J-R.Lee:94}.  Since the ordinary periodic boundary
conditions in that simulation corresponds to FBC's in spin models (cf.\ 
Sec.\ \Ref{Sec:Bg.bc}), whereas the present simulations are performed with
PBC's, the value of the cumulant at criticality is, however, not expected
to be the same\cite{Olsson:xyff}.

Figure \Ref{fig-binder}(b) shows the data collapse. Following Ref.\ 
\cite{J-R.Lee:94} we assume that $U_L(T) = \phi(tL^{1/\nu})$, where $t =
T/T_c - 1$. We then expand $\phi(x) = \phi_0 + \phi_1 x + \phi_2 x^2$ for
small $x$ and adjust these three parameters together with $\nu$ and $T_c$
to get the best possible fit. With data close to $T_c$, ($0.605 < U <
0.640$) for system sizes $L = 16$, 32, and 64, we obtain $T_c = 0.8244$ and
$\nu \approx 0.81$. This value of $\nu$ is in good agreement with the
published values, listed in the Introduction.

\newpage
\section{Correlation lengths}
\label{Sec:Corr}

In the previous Section we performed a number of analyses with methods that
take advantage of the finite size effects in the MC data.  This is usually
the most efficient way to determine the critical behavior from Monte Carlo
simulations.  The alternative approach is to determine the correlation
length from the length-dependence of some correlation functions and then
extract the critical behavior from its temperature dependence.  In the
present section we take this alternative route.

In order to test some techniques for determining the correlation length and
extracting the correlation length exponent and the critical temperature, we
first present an analysis of the 2D Ising model.  Because of the dual
advantage of a known critical behavior and a fast cluster update algorithm,
this model serves as a very convenient testing ground.  One minor
difference between the analysis of the 2D Ising model and the \FFXY\ model
is due to the anti ferromagnetic ordering.  Whereas the critical behavior
of the ferromagnetic Ising model is obtained in the $k\rightarrow0$ limit,
the corresponding critical behavior in the \FFXY\ model manifests itself at
$\k = (\pi,\pi)$. This is taken care of by performing all the analyses in
Sec.\ \Ref{Sec:Corr.xi} in terms of $\q = (\pi,\pi) - \k$ instead of $\k$,
cf.\ Sec.\ \Ref{Sec:Bg.dual-corr}.

In this section we determine two different characteristic lengths from our
MC data for $\left<v_\k v_{-\k}\right>$.  The reason that it is at all
possible to define \emph{two different} characteristic lengths is related
to the above discussion.  Whereas the screening length $\lambda$ is
determined from the $k\rightarrow 0$ limit of these correlations, the
correlation length $\xi$ is determined from the limit $q\rightarrow 0$.

For the determination of the characteristic lengths one like to have the
correlation function for an infinite system. It is therefore of great
importance to know when the undesired effects of the finite lattice size
set in.  Before applying the obtained techniques to the \FFXY\ model we do
a careful analysis of the finite-size effects in this model.  The results
corroborate the suggestion\cite{Olsson:xyff} that the correlation function
is plagued by finite size effects unless the system is large enough that
$\Upsilon\approx0$.

After these preliminaries we then turn to determinations of the correlation
length $\xi$, the critical exponent $\nu$, and the critical temperature
$T_c$, in the \FFXY\ model.  Much as expected from the evidence of two
distinct transitions, the behavior is found to be consistent with an
ordinary Ising transition, $\nu=1$.  However, for this demonstration it
turns out to be necessary to examine the behavior fairly close to $T_c$,
which corresponds to large correlation lengths, $\xi > 10$.  The screening
length $\lambda$ associated with the KT transition is also determined, and
its temperature dependence is found to be entirely different from the
behavior of $\xi$, but in good agreement with Kosterlitz'
result, Eq.\ (\Ref{l0.sqrt})\cite{Kosterlitz:74}.

We finally turn to the behavior of $\xi$ at $T < T_c$.  At these
temperatures we have no data with $\Upsilon\approx 0$, (and if we had,
there might also be problem fulfilling $L\gg\xi$) and we therefore need
some other methods to avoid the finite-size effects.  The solution is to
restrict the analysis to $T\leq\TKT$, where the different boundary
conditions of Sec.\ \Ref{Sec:Bg.bc} may be employed.  However, the
temperature dependence of $\xi$ below $T_c$ (and $\TKT$) does not seem to
be useful for assessing the critical behavior, possibly an effect of the
presence of the KT transition between the obtain data and $T_c$.

\subsection{The correlation length in the 2D Ising model}
\label{Sec:Corr.Ising}

The Hamiltonian of the Ising model is
\begin{displaymath}
  H^I = -J\sum_{\langle ij\rangle} s_i s_j,
\end{displaymath}
where $i$ and $j$ numerate the lattice points, $s_i = \pm 1$, and the
summation is restricted to nearest neighbors.  In two dimensions the
correlation length exponent is $\nu = 1$, and at a square lattice the
critical temperature is known to be
\begin{displaymath}
  T_c^I/J = \frac{2}{\ln\left(\sqrt{2} + 1\right)} \approx 2.269.
\end{displaymath}

Since $\nu = 1$, a plot of $1/\xi$ versus $T$ is expected to yield a
rectilinear behavior down to $T_c^I$. However, the verification of this
turns out to require data fairly close to $T_c^I$, large correlation
lengths, and therefore rather big lattices.  For most purposes this
exercise is pointless, since -- beside being obtained from the exact
solution -- the value $\nu = 1$ may be verified from MC simulations by
means of finite size scaling at $T_c^I$.  But since this kind of finite
size scaling does not seem to work in the \FFXY\ models for the accessible
lattice sizes, we have to resort to analyses of the correlation functions.
With that background, analyses of the correlation function for the 2D Ising
model serves as a help to develop techniques for similar analyses of the
\FFXY\ model.  Beside the benefit of the exactly known critical behavior,
the analysis of the Ising model is greatly simplified by means of the
cluster algorithm\cite{Wolff:89a} that is instrumental in obtaining MC with
small statistical errors.

\subsubsection{Determination of the correlation length}

At first sight the obvious way to determine the correlation length is to
examine the exponential decrease of the correlation function $g(r)$ down to
zero. This amounts to adjusting the parameters $A$ and $\xi$ to obtain best
possible fit to the expression,
\begin{displaymath}
  g(r) = A e^{-r/\xi}.
\end{displaymath}
At temperatures closely above $T_c$, the above expression
should be modified to take correlations across the whole system into
account.  This is customarily done by instead fitting to an expression with
the periodicity of the system,
\begin{equation}
  \label{g.r.fit}
  g(r) = A \left(e^{-r/\xi} + e^{-(L-r)/\xi}\right).
\end{equation}
It is, however, difficult to obtain reliable values for the correlation
length with this procedure. The main complication is that the optimum value
of $\xi$ does depend on the range in $r$ employed for the fit.  This is not
too surprising, since the pure exponential decay only is expected for very
small values of $g(r)$.

An alternative determination of $\xi$ by means of $g(k)$, the Fourier
components of the correlation function, has been suggested in Ref.\ 
\cite{Cooper_FP},
\begin{equation}
  \label{gk.xxi2}
  \xi = \frac{L}{2\pi} \sqrt{\frac{g(0)}{g(2\pi/L)} - 1}.
\end{equation}
An advantage with this expression is that no fitting is needed, and that the
arbitrariness involved in choosing the fitting interval is eliminated.
This expression may be derived from 
\begin{equation}
  \label{gk.xi}
  g(\k) \propto \frac{1}{\tk^2 + \xi^{-2}}.
\end{equation}
The relation to an exponential decay is obtained since the Fourier
transform of this function is the Bessel-$K_0$ function,
\begin{displaymath}
  g(r) \propto K_0(r/\xi),
\end{displaymath}
with the limiting behavior $\sim e^{-r/\xi}$.

\begin{figure}
  \hspace{2cm}\input{is.xi.xxi}\put(0,0){%
  \epsfig{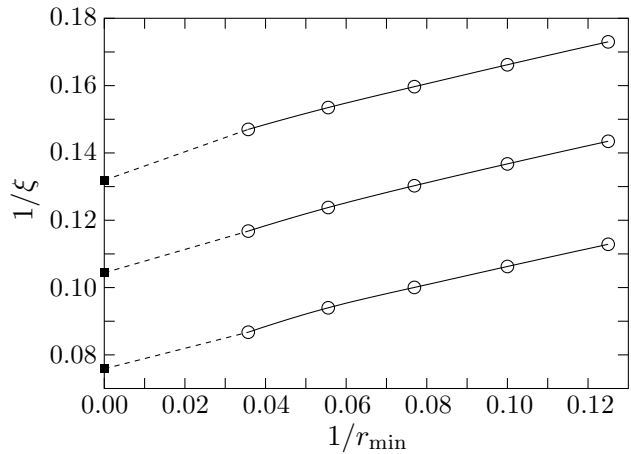}}\end{picture}
  \caption{\small Results for the correlation length in the 2D
  Ising model.  The open circles are $\xi$ from fitting $g(r)$ to Eq.\ 
  (\Ref{g.r.fit}) for $r_{\rm min} < r < 2 r_{\rm min}$. The solid squares
  are determinations of $\xi$ from the small-$k$ limit, Eq.\ 
  (\Ref{gk.xxi2}). The data obtained in that way thus seems to correspond
  to the $r_{\rm min} \rightarrow \infty$ limit. The MC data is for $L =
  128$ and, from top to bottom, $T/J = 2.45$, 2.41, and 2.37.}\label{fig-xi.xxi}
\end{figure}

In Fig.\ \Ref{fig-xi.xxi} we display some determinations of the correlation
length in the 2D Ising model.  The open circles are obtained with Eq.\ 
(\Ref{g.r.fit}) whereas the solid squares are from Eq.\ (\Ref{gk.xxi2}).
The curves are for different temperatures, from top to bottom, $T/J =
2.45$, 2.41, and 2.37.  The different values of $\xi$ are due to the
different ranges of $r$ for the data included in the fit. We use $g(r)$ for
$r_{\rm min} < r < 2 r_{\rm min}$. The $x$ axis shows $1/r_{\rm min}$.  We
find that as $r_{\rm min}$ increases, the inverse correlation length
decreases towards the solid squares from Eq.\ (\Ref{gk.xxi2}).  This
suggests that the $k\rightarrow 0$ limit really does probe the
long-distance limit.  On the basis of this comparison we believe that Eq.\ 
(\Ref{gk.xxi2}) gives a reliable way to determine the correlation length
$\xi$.

A precise determination of $\xi$ with Eq.\ (\Ref{gk.xxi2}) requires fairly
long MC simulations.  This is the case since $g(2\pi/L)$ and $g(0)$ measure
the amplitude of the largest fluctuations in the system, with the
correspondingly long decorrelation times.  A way to reduce the effect of
statistical errors is to include some more $k$-vectors in the analysis.
This is motivated by the difficulty to obtain good accuracy from Eq.\ 
(\Ref{gk.xxi2}) on data from the \FFXY\ model at large lattices ($L = 128$,
256).  However, using $g(k)$ in a too large interval will affect the
correlation length $\xi$.  Assuming that the exponential $r$-dependence
only holds for $r\gg \xi$, or, similarly, that the asymptotic
$k$-dependence only is valid for $k \ll 2\pi/\xi$ we restrict ourselves to
making use of data from wave-vectors $k<\pi/\xi$, only.  For small values
of $\xi$ this is not very restrictive, and since we are interested in the
small-$k$ limit we impose the additional condition $k < \sqrt{0.1}$.

The procedure to determine $\xi$ is then to first fit the data to
\begin{equation}
  \label{gk.g012}
  \frac{1}{g(\k)} = g_0 + g_1 \tk^2 + g_2 (\tk^2)^2,
\end{equation}
(where we also include a second order term in $\tk^2$ to take care of the
curvature in the data) and then extract the correlation length through
\begin{equation}
  \label{gk.kxi}
  \xi = \sqrt{g_1/g_0}.
\end{equation}
In the limiting case with only two wave-vectors, this procedure is readily
shown to be equivalent to Eq.\ (\Ref{gk.xxi2}).

\subsubsection{Determination of the critical behavior}

\begin{figure}
  \hspace{2cm}\input{is.kxi.t}\put(0,0){%
  \epsfig{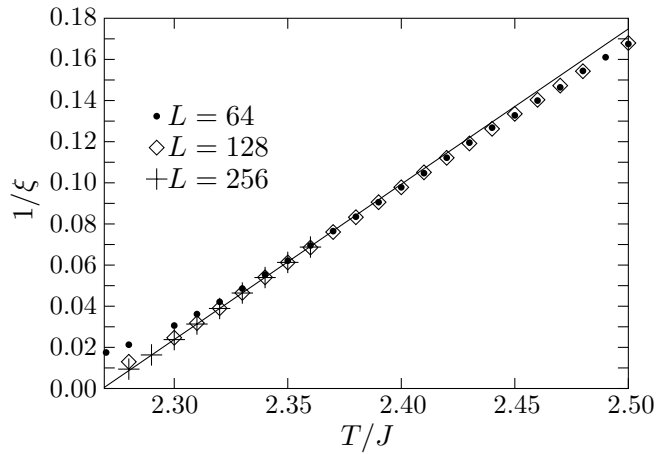}}\end{picture}
  \caption{\small Correlation length in the 2D Ising model for $L =
  64$, 128, and 256.  The solid line is from fitting to Eq.\ (\Ref{xi.AT})
  for $T/J \leq 2.33$ with data that fulfills $L/\xi > 5$. }\label{fig-is.kxi.t}
\end{figure}

Figure \Ref{fig-is.kxi.t} shows the temperature dependence of the
correlation length in the 2D Ising model for three different system sizes,
$L = 64$, 128, and 256. In the vicinity of $T_c^I$ the data reveals some
finite-size effects. The correlation length becomes smaller in a too small
system.  The present data seems to suggests that the determinations are
reliable only if $L/\xi > 5$.

\begin{figure}
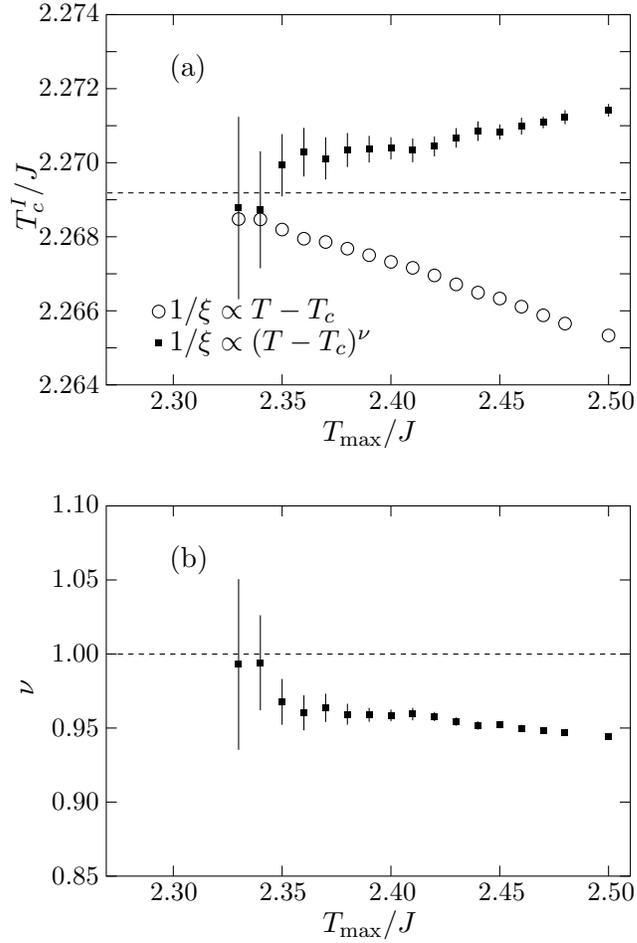

  \hspace{2cm}\input{is.tc.tmax}\put(0,0){%
  \epsfig{file=is.tc.tmax.eps}}\end{picture}
  \par\vspace{0.5cm}\noindent
  \hspace{2cm}\input{is.nu.tmax}\put(0,0){%
  \epsfig{file=is.nu.tmax.eps}}\end{picture}
  \caption{\small Determinations of $T_c$
  and $\nu$ in the 2D Ising model from the correlation length data by
  fitting to Eqs.\ (\Ref{xi.AT}) and (\Ref{xi.ATnu}).  The open circles are
  from fits with the exponent keeped fixed, $\nu = 1$, whereas the solid
  squares are obtained with both $T_c$ and $\nu$ as free parameters.  The
  general trend is that both quantities approach the exact values,
  indicated by the dashed lines, as the temperature interval is
  decreased.}\label{fig-is.tc.tmax}
\end{figure}

Also apparent in the figure is a slight curvature in the data. The
expected linear behavior is found only right above $T_c^I$. Estimates of
$T_c^I$ may be obtained by fitting 
\begin{equation}
  1/\xi \propto A (T - T_c),
  \label{xi.AT}
\end{equation}
with $A$ and $T_c$ as free parameters. In these analyses we only make use
of data for $L = 128$ and 256.  Due to the curvature in the data, the
critical temperature obtained in this way does depend on the temperature
interval for the fit.  For temperatures closely above $T_c^I$ we only
include data points with $L/\xi > 5$. This gives $L$-dependent lower limits
for the temperature interval. The upper limit of the temperature interval
is given by $T_{\rm max}$.  The fit is then performed for several different
values of $T_{\rm max}$. The dependence of $T_c^I$ on $T_{\rm max}$ is
shown by open circles in Fig.\ \Ref{fig-is.tc.tmax}(a). The dashed line is
the exact value of $T_c^I$. For large $T_{\rm max}$ (large temperature
intervals) the analysis yields too low estimates of the critical
temperature, but with decreasing $T_{\rm max}$ the estimated $T_c^I$
increases towards the correct value.

This linear fit presumes a known value of the correlation length exponent
$\nu$. Since the value of $\nu$ in the \FFXY\ model is highly disputed it
is also of interest to perform the fit with $\nu$ as a free parameter.
That is done by fitting
\begin{equation}
   1/\xi = A (T - T_c)^\nu,
  \label{xi.ATnu}
\end{equation}
with $\nu$, $A$ and $T_c$, as free parameters. Again we repeat the
analysis for several different $T_{\rm max}$.  The values of $T_c^I$ and
$\nu$ as functions of $T_{\rm max}$ are shown by solid squares in Figs.\ 
\Ref{fig-is.tc.tmax}. We find that both quantities
approach the expected values as the temperature interval is reduced. The
erratic behavior at low values of $T_{\rm max}$ is due to statistical
errors that become significant in such narrow temperature intervals.

\subsection{Finite size effects}

In the previous section we found, much as expected, that it is necessary to
perform the MC simulations of the 2D Ising model at systems that are
considerably larger than the correlation length.  The criterion was $L/\xi
> 5$.  When finite size effects set in we get larger correlations and,
thereby, too small correlation length.  In this section we demonstrate that
in the analysis of the \FFXY\ model, this condition has to be supplemented
by a second one that is related to the additional $XY$ degrees of freedom,
and the corresponding screening length, $\lambda$.  In terms of the
helicity modulus this condition may be written $\Upsilon\approx 0$, which
is equivalent to $L\gg\lambda$.

\begin{figure}
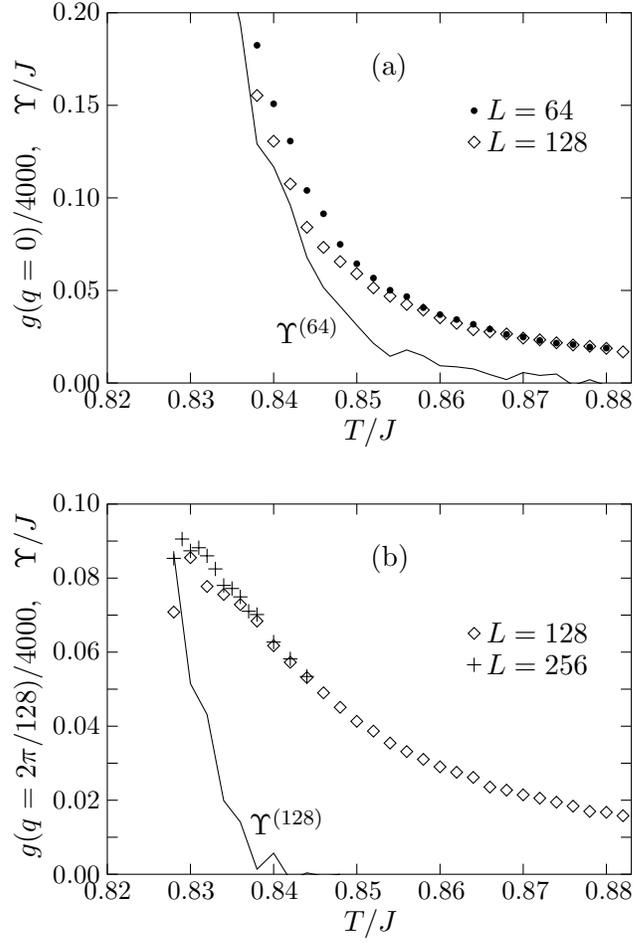

  \hspace{2cm}\input{finite.64}\put(0,0){%
  \epsfig{file=finite.64.eps}}\end{picture}
  \par\vspace{0.5cm}\noindent
  \hspace{2cm}\input{finite.128}\put(0,0){%
  \epsfig{file=finite.128.eps}}\end{picture}
  \caption{\small The correlation
  function $g(q)$ for two different lattice sizes, together with the
  helicity modulus for the smaller size.  Panel (a) is the $q=0$ component
  for system sizes $L = 64$ and 128 whereas panel (b) is the $q=2\pi/128$
  component for $L = 128$ and 256. In both cases $g(q)$ start to differ
  when $\Upsilon\neq0$.}\label{fig-finite}
\end{figure}

In order to check where the finite size effects become important for the
determination of $\xi$, it is convenient to monitor $g(q)$ for $q \approx
0$ from different lattice sizes as functions of temperature. This is done
in Fig. \Ref{fig-finite}.  Panel (a) shows $g(q = 0)$ for lattice sizes $L =
64$ and 128.  $g(q=0)$ for the even larger system, $L = 256$, suffers from
large statistical errors and is therefore not included in the figure.
Comparison between $L = 128$ and 256, are instead performed with $g(q =
2\pi/128)$, shown in Fig.\ \Ref{fig-finite}(b).

In both these figures the helicity modulus for the smaller size is
included as solid lines. Comparing the data for the different sizes we find
that they start to differ at about the temperature where the helicity
modulus becomes appreciably different from zero.  This implies that
determinations of the correlation length from $g(q)$ are uncertain if the
system is not large enough to ensure that $\Upsilon\approx 0$.

\begin{figure}
  \hspace{2cm}\input{epsk840}\put(0,0){%
  \epsfig{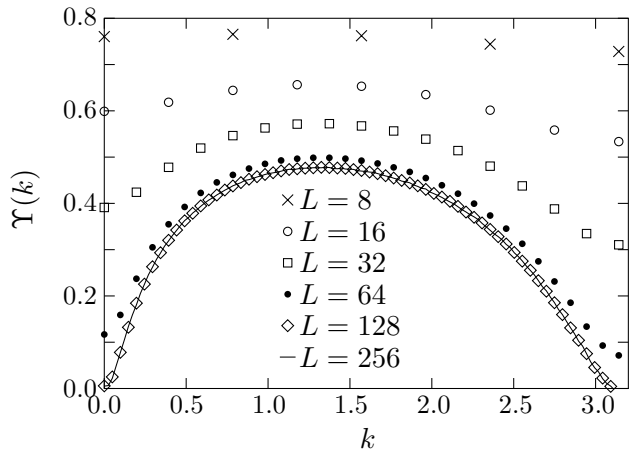}}\end{picture}
  \caption{\small Size dependence of $\Upsilon(\k)$ at $T/J
  = 0.840$ for $\k = (k_x,0)$.  The message in this figure is that
  $\Upsilon(\k)$ and thereby $g(\k)$ is size-dependent unless $\Upsilon(\k
  = 0) = 0$.}\label{fig-epsk840}
\end{figure}

The suggested link from the two previous figures between the
size-dependence in $g(q)$ and the helicity modulus may be understood by
examining the wave-vector dependent helicity modulus $\Upsilon(k)$ for
different system sizes.  Implicit in this discussion is the close relation
between $g(\k)$ and $\Upsilon(\k)$ in Eqs.\ (\Ref{gk.mk}) and
(\Ref{def.Upsk.m}).  Figure \Ref{fig-epsk840} shows $\Upsilon(k)$ at $T/J =
0.84$, which is well above both $\TKT$ and $T_c$.

At each temperature above $\TKT$ the helicity modulus $\Upsilon \equiv
\Upsilon(k=0)$ vanishes for sufficiently large $L$.  A finite value of
$\Upsilon$ may therefore be considered a finite-size effect.  The message
of Fig.\ \Ref{fig-epsk840} is that if $\Upsilon$ suffers from finite-size
effects, then the same is true for all the other components $\Upsilon(k)$,
as well.  The converse also appears to be true.  For system sizes with
$\Upsilon \approx 0$, $\Upsilon(k)$ is independent of $L$. This is
illustrated by the two largest systems $L = 128$ (diamonds) and 256 (solid
line), with data just on top of each other.

From the relations between $\Upsilon(\k)$ and $g(\k)$ the above result is
of relevance for $g(\k)$.  We therefore conclude that our results for the
correlation function are without significant finite size effects only if
$\Upsilon\approx0$, which means that precisely this condition has to be
fulfilled to facilitate reliable determinations of the correlation length
$\xi$.

In the following determination of $T_c$ and $\nu$ we only make use of the
data for the larger lattices, $L = 128$ and 256.  For $L = 128$ the
criterion $\Upsilon \approx 0$ suggests making use of data for $T/J >
0.84$, only.  The corresponding temperature limit for $L = 256$ is more
difficult to obtain since $\Upsilon^{(256)}$ suffers from large statistical
errors.  The temperature limit we have used, $T > 0.828 J$, is obtained
from considering the fraction $L/\lambda$.  The temperature limit for $L =
128$ gives $(L/\lambda)_{\rm min} > 27$, which for $L = 256$ is fulfilled
only for $T/J > 0.828$.  The determination of $\lambda$ is discussed in
Sec.\ \Ref{Sec:Corr.lambda}.

One should, of course, keep in mind that the usual finite size effect,
related to the fraction $L/\xi$, also may be relevant in these systems. It
does, however, seem that this condition is the more restrictive one only
for sizes $L>256$.

\subsection{The correlation length in the \FFXY\ model}
\label{Sec:Corr.xi}

We now apply the methods and results from the previous Sections to our MC
data for the \FFXY\ model.

\begin{figure}
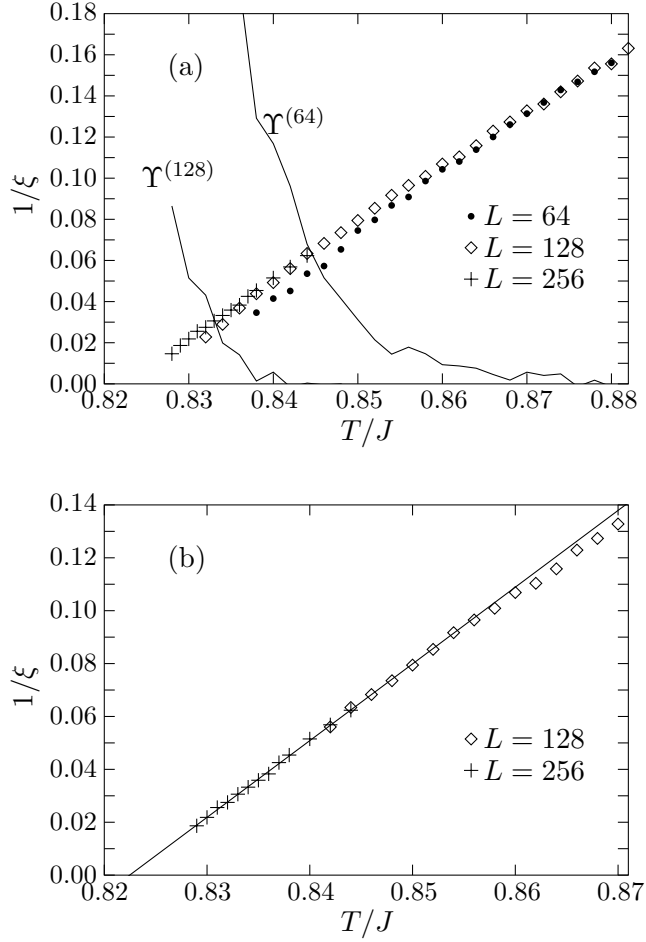

  \hspace{2cm}\input{kxi.Ups.t}\put(0,0){%
  \epsfig{file=kxi.Ups.t.eps}}\end{picture}
  \par\vspace{0.5cm}\noindent
  \hspace{2cm}\input{kxi.t}\put(0,0){%
  \epsfig{file=kxi.t.eps}}\end{picture}
  \caption{\small Correlation length
  $\xi$ in the \FFXY\ model.  (a) Examination of the finite size effects.
  The data corroborates the conclusion that the determinations of $\xi$
  suffer from finite size effects if $\Upsilon\neq0$.  (b) The correlation
  length data to be used in the determinations of $T_c$ and $\nu$. The
  solid line is obtained by fitting to Eq.\ (\Ref{xi.AT}) for $T \leq
  0.842J$.  Estimates of the statistical uncertainities in $\xi$ are found
  in Table \Ref{tab:Corr}.}\label{fig-kxi.t}
\end{figure}

Figure \Ref{fig-kxi.t}(a) shows our values for $\xi$ obtained by
self-consistently fitting $g(\q)$ with $q < \pi/\xi$ and $q < \sqrt{0.1}$
to Eq.\ (\Ref{gk.g012}) for systems of size $L = 64$, 128, and 256. The
data for the larger sizes are also given in Table \Ref{tab:Corr}. Also
shown is the helicity modulus for the two smaller sizes. We note that $\xi$
for $L=64$ (solid dots) start to deviate from the results for the larger
lattice ($L=128$, open squares) at the temperature where $\Upsilon^{(64)}$
becomes appreciably different from zero.  The corresponding situation holds
for $\xi$ obtained with $L=128$ and 256.  Note also that this finite
size effect is somewhat peculiar since analysis of data from a smaller
lattice yields a \emph{too large} value of the correlation length.  This is
opposite to the usual case, cf.\ Fig.\ \Ref{fig-is.kxi.t}.

After skipping the data affected by finite size effects, the analysis of
the correlation function in the \FFXY\ model, becomes very similar to the
corresponding analysis of the 2D Ising model in Sec.\ \Ref{Sec:Corr.Ising}.
The data to be used is shown in Fig.\ \Ref{fig-kxi.t}(b).

\begin{figure}
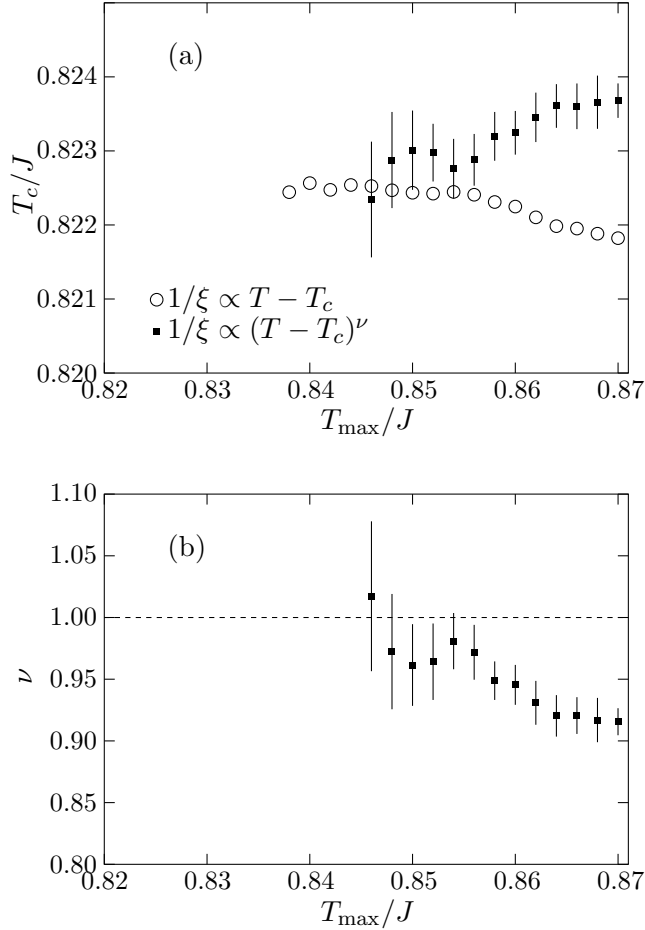

  \hspace{2cm}\input{tc.tmax}\put(0,0){%
  \epsfig{file=tc.tmax.eps}}\end{picture}
  \par\vspace{0.5cm}\noindent
  \hspace{2cm}\input{nu.tmax}\put(0,0){%
  \epsfig{file=nu.tmax.eps}}\end{picture}
  \caption{\small Determinations of
  $T_c$ and $\nu$ from the temperature dependence of $\xi$.  This is a
  careful analysis of the data shown in Fig.\ \Ref{fig-kxi.t}(b) and listed
  in Table \Ref{tab:Corr}.  The fits are performed with several different
  temperature intervals, including data up to $T_{\rm max}$.  The open
  circles are obtained by fitting to Eq.\ (\Ref{xi.AT}), i.e.\ assuming
  that $\nu = 1$. The solid squares are obtained with $\nu$ as a free
  parameter by fitting to Eq.\ (\Ref{xi.ATnu}).  Panel (a) shows the
  obtained critical temperatures whereas panel (b) show the corresponding
  values of $\nu$.  This is very similar to the behavior in the 2D Ising
  model, shown in Fig.\ \Ref{fig-is.tc.tmax}.}\label{fig-tc.tmax}
\end{figure}

For determinations of $T_c$ we first assume $\nu = 1$ and fit our data to
Eq.\ (\Ref{xi.AT}) for temperatures $T \leq T_{\rm max}$.  Our results for
the critical temperature vs.\ $T_{\rm max}$, are shown as open circles in
Fig.\ \Ref{fig-tc.tmax}(a).  The figure shows a slowly increasing trend in
$T_c$ for decreasing $T_{\rm max}$, similar to the results for the Ising
model, Fig.\ \Ref{fig-is.tc.tmax}(a).  From this we get our best value of the
$Z_2$ temperature, $T_c/J \approx 0.8225(5)$.  The line in Fig.\
\Ref{fig-kxi.t}(b) is from the fit with $T_{\rm max}/J = 0.842$. 

The next step is to do a similar fit with $\nu$ as a free parameter, by
fitting to Eq.\ (\Ref{xi.ATnu}). Our values of $T_c$ and $\nu$ are shown by
solid squares as functions of $T_{\rm max}$ in Fig.\ \Ref{fig-tc.tmax}.
For large $T_{\rm max}$ the analysis gives non-Ising exponents, $\nu
\approx 0.9$, but as $T_{\rm max}$ decreases (the temperature interval
shrinks) the data suggest an increasing trend in $\nu$ towards 1.0.  From
this analysis of the temperature dependence of $\xi$ we therefore conclude
that the data, indeed, is consistent with $\nu = 1$.

We also note that this dependence of $\nu$ on $T_{\rm max}$ is very similar
to the corresponding analysis in the 2D Ising case, Fig.\ 
\Ref{fig-is.tc.tmax}(b).  Analyses based on data from a somewhat too large
temperature interval suggested $\nu < 1.0$, but with decreasing temperature
interval the correct value of the exponent was obtained.  This comparison
serves to strenghten our conclusion of the ordinary 2D Ising value for the
correlation length exponent in the \FFXY\ model.

\begin{table}[htbp]
  \begin{center}
    \leavevmode
    \begin{tabular}[l]{ccccc}
      $T/J$ & $\xi^{(128)}$ & NMCS$/10^6$ & $\xi^{(256)}$ &
      NMCS$/10^6$ \\
      \hline
       0.828 &              &       &    68.2(2.4) &   23 \\ \cline{4-5}
       0.829 &              &       &    53.6(1.5) &   23 \\
       0.830 &              &       &    45.8(1.5) &   22 \\
       0.831 &              &       &    39.1(1.0) &   18 \\
       0.832 &    43.9(1.9) &  12   &    36.3(9)   &   18 \\
       0.833 &              &       &    32.6(8)   &   18 \\
       0.834 &    34.6(1.3) &   8   &    30.0(6)   &   18 \\
       0.835 &              &       &    27.9(5)   &   18 \\
       0.836 &    27.1(5)   &  40   &    26.1(4)   &   18 \\
       0.837 &              &       &    23.5(3)   &   18 \\
       0.838 &    22.8(3)   &  58   &    22.0(3)   &   17 \\
       0.840 &    20.3(3)   &  56   &    19.4(2)   &   18 \\ \cline{2-3}
       0.842 &    17.8(2)   &  56   &    17.6(2)   &   19 \\
       0.844 &    15.78(11) &  92   &    16.0(2)   &   19 \\
       0.846 &    14.64(12) &  64   &              &      \\
       0.848 &    13.60(9)  &  64   &              &      \\
       0.850 &    12.59(8)  &  64   &              &      \\
       0.852 &    11.71(9)  &  32   &              &      \\
       0.854 &    10.91(7)  &  40   &              &      \\
       0.856 &    10.36(6)  &  40   &              &      \\
       0.858 &    9.91(6)   &  40   &              &      \\
       0.860 &    9.36(9)   &  16   &              &      \\
       0.862 &    9.06(8)   &  16   &              &      \\
       0.864 &    8.63(8)   &  16   &              &      \\
       0.866 &    8.14(10)  &   9   &              &      \\
       0.868 &    7.85(7)   &  16   &              &      \\
       0.870 &    7.53(8)   &  10   &              &      \\
       \hline
    \end{tabular}
    \caption{Correlation length versus temperature for system sizes $L =
      128$ and 256, together with rough estimates of the associated
      statistical errors.  NMCS is the number of Monte Carlo sweeps through
      the system for the respective sizes.  The data above the horizontal
      lines suffer from finite size effects and are therefore not used in
      the determinations of $T_c$ and $\nu$ in Sec.\ \Ref{Sec:Corr.xi}.}
    \label{tab:Corr}
  \end{center}
\end{table}

\subsection{Screening length $\lambda$}
\label{Sec:Corr.lambda}

\begin{figure}
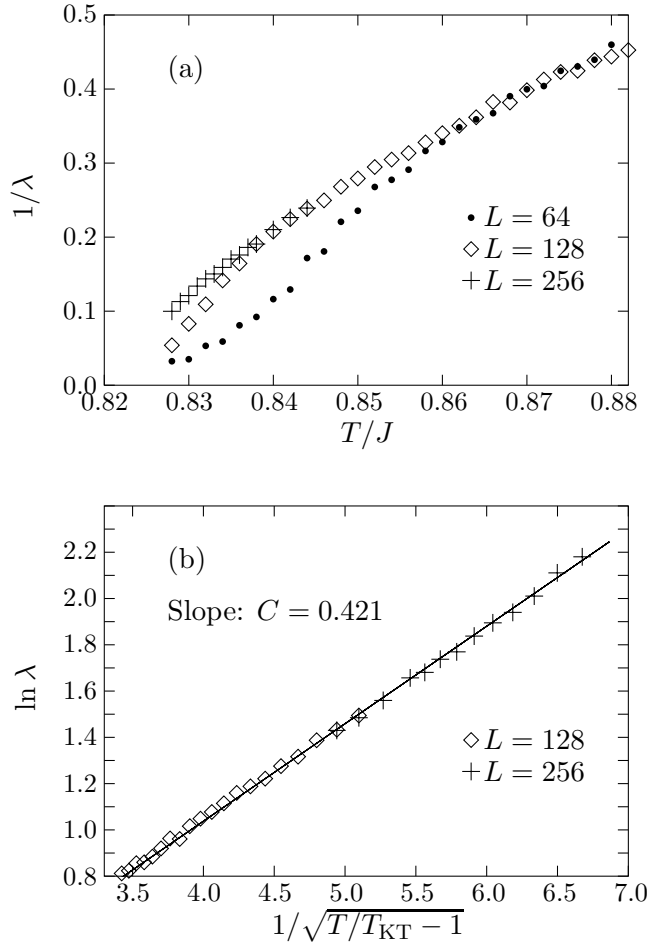

  \hspace{2cm}\input{lambda}\put(0,0){%
  \epsfig{file=lambda.eps}}\end{picture}
  \par\vspace{0.5cm}\noindent
  \hspace{2cm}\input{ln.lambda}\put(0,0){%
  \epsfig{file=ln.lambda.eps}}\end{picture}
  \caption{\small Temperature
  dependence of the screening length, $\lambda$.  (a) The inverse screening
  length versus temperature.  The finite size effect gives to large values
  of $\lambda$.  Considering the values from the largest lattices (the
  uppermost values) the data has a clear curvature.  This is in contrast to
  the corresponding behavior of $1/\xi$ in Fig.\ \Ref{fig-kxi.t}.  This is
  a clear demonstration that $\lambda$ behaves very differently from $\xi$.
  (b) Verification of the Kosterlitz temperature dependence for $\lambda$.
  Here $\TKT/J = 0.8108$ from Sec.\ \Ref{Sec:Finite.TKT} and the data
  affected by finite size effects is removed.  The slope from this figure
  is, however, not in accordance with $C = 0.54\pm 0.02$ from the
  finite-size scaling analysis in Sec.\ \Ref{Sec:Finite.TKT}.}\label{fig-lambda}
\end{figure}

It is also of great interest to determine the screening length $\lambda$,
associated with the free vortices in the system. This turns out to, in some
respects, be similar to the above determination of $\xi$.  The starting
point is the expected behavior for the wave-vector dependent helicity
modulus\cite{Minnhagen:review},
\begin{displaymath}
  \Upsilon(\k) \equiv \frac{J}{\epsilon(\k)}
  = \frac{J}{\tilde{\epsilon}} \frac{\tk^2}{\tk^2 + \lambda^{-2}},
\end{displaymath}
where $\lambda$ is the screening length associated with free vortices and
$\tilde{\epsilon}$ is due to the polarization of bound
pairs\cite{Minnhagen:review}.  Neglecting the $k$-dependence in
$\tilde{\epsilon}$ one expects $\tk^2/ \Upsilon(k)$ to be linear in
$\tk^2$. This is, however, not quite the case.  The data shows a negative
curvature and we therefore perform a fit to the second-order polynomial
\begin{displaymath}
  \frac{\tk^2}{\Upsilon(k)} = a_0 + a_1 \tk^2 + a_2 (\tk^2)^2.
\end{displaymath}
Assuming that the above equation holds for wavelengths larger than
$\lambda$ we, self-consistently, make use of data for $k<\pi/\lambda$,
only. The screening length is then obtained from $\lambda =
\sqrt{a_1/a_0}$. The results are shown in Fig.\ \Ref{fig-lambda}(a).  Note,
again that the finite size effects set in for $L=64$ and 128, at
temperatures below $\approx 0.87$ and $\approx 0.84$, respectively.

We also note that the data obtained does fit well to the well-known
Kosterlitz expression for the characteristic length\cite{Kosterlitz:74}.
This is shown in Fig.\ \Ref{fig-lambda}(b) where we plot $\ln\lambda$ vs.\ 
$1/\sqrt{T/\TKT-1}$.  From this linear curve it is possible to get a value
for the screening length at $T_c$.  An extrapolation to $T_c/J = 0.8225$
gives $\lambda(T_c) \approx 17.7$.  From our rough criterion for negligable
finite size effects in Sec.\ \Ref{Sec:Corr.xi}, $L/\lambda > 27$, we may
then estimate that the finite size effects associated with the KT
transition would be negligable for systems with $L > 478$.

The slope $C \approx 0.42$ is not quite in accordance with $C\approx 0.54$
obtained from the finite-size analysis of $\Upsilon_L$ in Sec.\ 
\Ref{Sec:Finite.TKT}.  This kind of difference was also found in the
ordinary $XY$ model with no frustration\cite{Olsson:Kost-fit}. However, it
is only the analysis in Sec.\ \Ref{Sec:Finite.TKT} that probes the region
immediately around $\TKT$, and this appears therefore to be the more
reliable one when it comes to determining the asymptotic behavior.

\subsection{Different boundary conditions}
\label{Sec:Corr.bc}

\begin{figure}
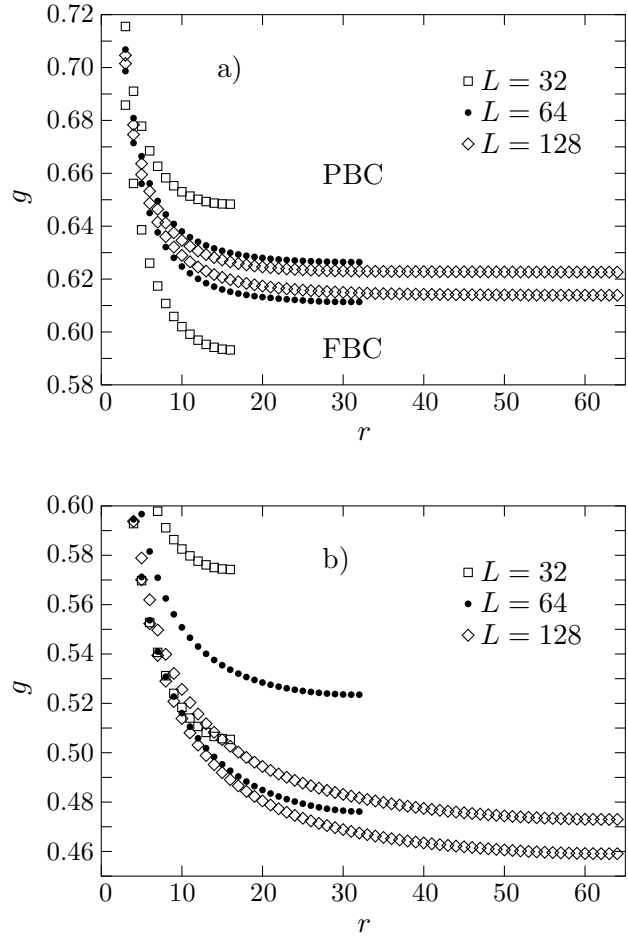

  \hspace{2cm}\input{tw.810}\put(0,0){%
  \epsfig{file=tw.810.eps}}\end{picture}
  \par\vspace{0.5cm}\noindent
  \hspace{2cm}\input{tw.820}\put(0,0){%
  \epsfig{file=tw.820.eps}}\end{picture}
  \caption{\small The correlation function
  $g(r)$ with different boundary conditions.  Panel (a) is for $T = 0.81J
  \approx\TKT$, whereas panel (b) is for $\TKT < T = 0.82J < T_c$.  At the
  lower temperature the correlation functions obtained with the two
  different boundary conditions scale with the system size in opposite
  ways, which facilitates a determination of the behavior in the
  thermodynamic limit.  At the higher temperature this is no longer true
  and there is no easy way to extrapolate to the behavior in the
  thermodynamic limit.}\label{fig-810.820}
\end{figure}

Figure \Ref{fig-810.820} show $g(r)$ versus $r$ at $T/J = 0.81$, and 0.82,
respectively, obtained with both PBC's and FBC's for lattice sizes $L =
32$, 64, and 128.  At the lower temperature the results from larger
lattices are squeezed in between the PBC and FBC results for a smaller
system.  When this behavior is valid, it seems safe to conclude that the
behavior of an infinite system is somewhere between these two limits, and,
to an excellent approximation, may be obtained as the average of these two
curves.

The behavior is dramatically different at the higher temperature.  This is
the case even though this temperature is well below $T_c$. (We expect this
different behavior to set in for $T>\TKT$.)  At this higher temperature,
$g(r)$ for both PBC's and FBC's decrease with increasing lattice size, and
it is therefore not possible to obtain any safe estimate for the behavior
in the thermodynamic limit.

\begin{figure}
  \hspace{2cm}\input{tw.M.t}\put(0,0){%
  \epsfig{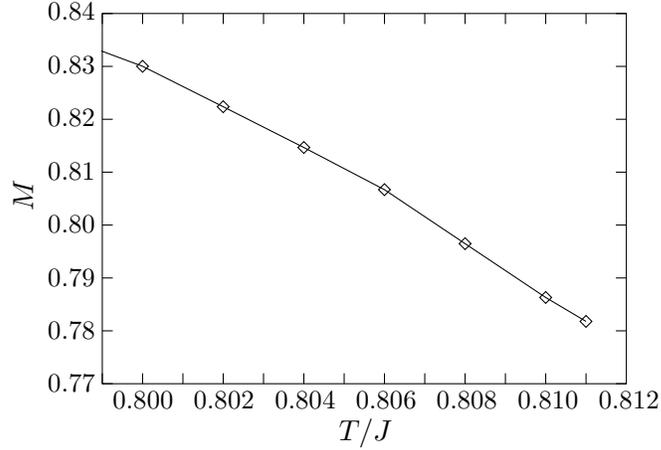}}\end{picture}
  \caption{\small The staggered magnetization.  The figure shows
  the average of values obtained with PBC's and FBC's for $L = 128$.  Since
  these sets of values are very close (difference $< 0.01$) we expect this
  average to be an excellent approximation of the behavior in the
  thermodynamic limit.}\label{fig-M.t}
\end{figure}

We are now in the position to determine $M_\infty(\TKT)$.  This is of great
interest since our argument for two distinct transitions in section
\Ref{Sec:Finite.univ.two} was based on a simple way to estimate this
quantity.  Our values for $M_\infty(T)$ are obtained by taking the average
between data from PBC's and FBC's.  $M_\infty$ obtained in this way is
shown in Fig.\ \Ref{fig-M.t}.  With $\TKT/J = 0.8108$ we find
$M_\infty(\TKT) \approx 0.783(2)$, which, indeed, is a good candidate to
the $L \rightarrow \infty$ limit in Fig.\ \Ref{fig-m.L}.  The results from
our more elaborated analyses are thus in very good agreement with the
simple approach of Sec.\ \Ref{Sec:Finite.univ.two}.

\subsection{The correlation length $\xi$ for $T < T_c$}

In this section we focus on the $Z_2$ correlation length in the
low-temperature region, $T < T_c$.  As discussed above we actually need
$g(q)$ for an infinite system for a reliable determination of $\xi$. As
shown in Fig.\ \Ref{fig-810.820}(a) the finite size effects may
be virtually eliminated by taking the average of $g(r)$ for PBC's and
FBC's.  The same turns out to be true for the Fourier components $g(q)$.
However, as discussed above this only works at $T < \TKT$, which means that
it does not seem possible to get any values for $\xi$ right below $T_c$.

The starting point for the present analysis is MC data obtained with both
PBC's and FBC's for a lattice of size $L = 128$, in a temperature interval
$0.770 \leq T/J \leq 0.811$. Much as in Sec.\ \Ref{Sec:Corr.xi} we fit
$1/g(q)$ to an expansion in $\tq^2$, cf.\ Eq.\ (\Ref{gk.g012}).  The data
should be taken at small $q$; we restrict the analysis to $qq <
\sqrt{0.1}$.  The main difference compared to the high-temperature case, is
that the data for $q=0$ has to be excluded at low temperatures. This is so
since $g(q=0)$ is directly related to the staggered magnetization squared
and $g(q)$ therefore is not a smooth function at $q=0$, in the low
temperature phase.

\begin{figure}
  \hspace{2cm}\input{tw.xi.tlow}\put(0,0){%
  \epsfig{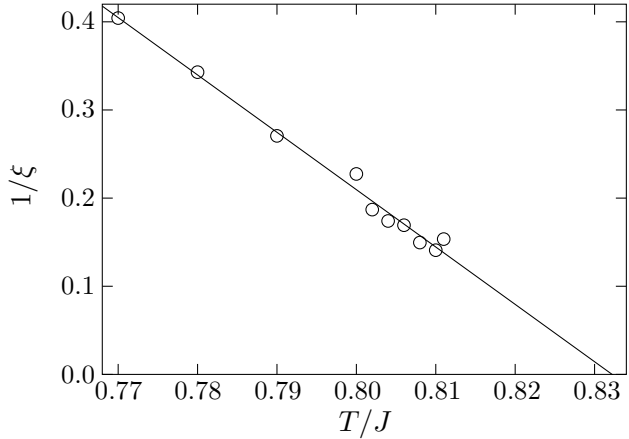}}\end{picture}
  \caption{\small The $Z_2$ correlation length $\xi$ in
  the temperature region below $T_c$ (and $\TKT$). Because of the dramatic
  change in the system at the KT transition, the critical temperature may
  not be determined from this kind of data.}\label{fig-xi.tlow}
\end{figure}

Figure \Ref{fig-xi.tlow} shows the obtained values of $\xi$. There are two
things to note. First, the data fall to a good approximation on a straight
line, $1/\xi \propto (T - T^*)$, but with the temperature $T^*$
significantly different from $T_c \approx 0.8225J$. The reason for this is
possibly that $\TKT$ lies between this data and $T_c$, and the
system undergoes a dramatic change at the KT transition.

The second point of interest is the value of the correlation length at
$\TKT$.  The finite size scaling analysis of $\Upsilon$ in Sec.\ 
\Ref{Sec:Finite.TKT} would not be reliable if there were significant finite
size effects associated with the $Z_2$ degrees of freedom.  We therefore
need data with $L \gg \xi$.  From Fig.\ \Ref{fig-xi.tlow} we find
$\xi(\TKT) \approx 7.2$ which means that the systems used in the finite
size scaling analysis, indeed, are considerably larger than the correlation
length; $L/\xi \geq 4.4$ for the smallest system, $L = 32$.

\section{Discussion}
\label{Sec:Disc}

Our conclusion for the \FFXY\ models with Villain or
cosines\cite{Olsson:xyff} interaction is thus that there are two distinct
transitions, with $\TKT < T_c$.  This is in contrast to the conclusions from
a study of frustrated systems with a variable bond strength $-\eta J$ at
one bond per plaquette\cite{Berge_DGL}.  Here $\eta = 1$ corresponds to the
fully frustrated case, and both smaller and larger values of $\eta$ were
used in order to separate the transitions, with the hope that this
information would shed light on the behavior of the \FFXY\ model, $\eta =
1$.  The conclusions -- Fig.\ 7 in Ref.\ \cite{Berge_DGL} -- were that as
$\eta$ increases above $\eta_c = 1/3$ there are two distinct transitions
-- with $T_c < \TKT$ -- which merge (or nearly so) at $\eta = 1$.  As
$\eta$ is increased further the transitions separate again, still with
the Ising-like transition at the lower temperature.

Since the result at $\eta = 1$ were not precise enough to exclude two
distinct transitions, it is perfectly possible to harmonize these results
with our present finding ($\TKT < T_c$ for $\eta = 1$).  The resulting
picture is then the following.  As a function of $\eta$ the temperatures
$\TKT$ and $T_c$ approach each other, cross at $\eta$ below but close to 1,
cross again above but close to $\eta = 1$ and then separate as $\eta$
increases.  While this might be an unexpected scenario, it seems to be
consistent with the conclusions based on the $XY$ Ising model in Ref.\ 
\cite{L_Granato_K}.

To summarize the findings of the present paper, we have found ample
evidence for two distinct transitions with $\TKT < T_c$ in the Villain
version of the \FFXY\ model. Our results corroborate the conclusions
presented in Ref.\ \cite{Olsson:xyff}.  We have given a strong argument for
two distinct transitions based on the universal jump condition alone, which
amounts to demonstrating that the staggered magnetization is finite, and
actually quite large, at the KT transition.  Since this result is obtained
with no complicated analysis or detailed fitting whatsoever, we consider
this to be a very robust argument for the existence of two transitions in
the \FFXY\ model.

A detailed finite size scaling analysis of the helicity modulus gives at
hand that the model undergoes an ordinary KT transition at $\TKT/J =
0.8108(1)$, and also that the size-dependence of this quantity in narrow
regions both below and above $\TKT$ is just as expected for a KT
transition.  Furthermore, with this value for $\TKT$ it is found that the
temperature dependence of the screening length is in accordance with the
well-known Kosterlitz' expression.

From thorough studies of the correlation functions in both the \FFXY\ model
and the 2D Ising model, we have determined $\xi$ and the correlation length
exponent $\nu$.  Our studies give at hand that the obtained values of $\nu$
depend on the size of the temperature intervals from which $\xi$ are taken,
and that $\nu$ approaches 1.0 as the size of the temperature interval is
decreased.  The result is therefore clearly in favor of ordinary Ising
exponents in the \FFXY\ model.  Using $\nu = 1$ the $Z_2$ critical
temperature was determined to $T_c/J = 0.8225(5)$.

\section*{Acknowledgements}
The author thanks Prof.\ Steve Teitel for suggesting this investigation and
both him and Prof.\ Petter Minnhagen for critical reading of the
manuscript.  Support from the Swedish Natural Research Council on contract
No.\ E-EG 10376-303 is gratefully acknowledged.

\newpage
\appendix
\section{Duality transformation for a frustrated model}
\label{App:dual}

The exact duality transformation\cite{Jose_KKN} of the $XY$ model with
Villain interaction is between the $XY$ model with PBC's and the Coulomb
gas with a term proportional to the polarization
squared\cite{Vallat_Beck,Olsson:self-cons.long}.  The derivation and
notations below closely follows Ref.\ \cite{Olsson:self-cons.long}.  For an
arbitrary set $\{A_{ij}\}$ the partition function is
\begin{equation}
  Z = \int_{-\pi}^\pi \prod_l \frac{d\theta_l}{2\pi}
  \exp\left(-\beta\sum_{\left<ij \right>} U(\theta_i - \theta_j - A_{ij})
  \right).
  \label{Zstart}
\end{equation}
With the Fourier expansion of the Boltzmann factor
\begin{displaymath}
  e^{-\beta U(\phi)} = \sum_{h = -\infty}^\infty \frac{e^{-h^2/2\beta
  J}}{\sqrt{2\pi\beta J}} e^{ih\phi},
\end{displaymath}
the partition function becomes
\begin{displaymath}
  Z = \sum_{\{h_{ij}\}=-\infty}^{\infty}
  \exp\left(i\sum_{\left<ij \right>} h_{ij} A_{ij} \right)
  \int \prod_l \frac{d\theta_l}{2\pi}
  \exp\left(i\theta_l \sum_{j\in\left<jl\right>} h_{jl}\right)
  \prod_{\left<ij\right>} \frac{e^{-h_{ij}^2/2\beta J}}{\sqrt{2\pi\beta J}}.
  \label{Z.h.theta}
\end{displaymath}
Integration over $\theta_l$ gives restrictions on the set $\{h_{ij}\}$,
$\sum_j h_{jl} = 0$ for each $l$.  A new set of variables, defined at the
centers of the plaquettes, take care of these restrictions:
\begin{eqnarray}
  h_\r^x & = & S_{\r+\yhat/2} - S_{\r-\yhat/2}, \nonumber\\
  h_\r^y & = & S_{\r-\xhat/2} - S_{\r+\xhat/2},
  \label{hS}
\end{eqnarray}
together with $\S = (S_x, S_y)$ which are included to allow for $\sum_\r
h_\r^\mu \neq 0$, $\mu = x$, $y$\cite{Olsson:self-cons.long}.  To take care
of the frustration we make use of $\D \times \A_\r = 2\pi f_\r$.
\begin{eqnarray*}
  \sum_\r h^x_\r A^x_\r + h^y_\r A^y_\r & = & \sum_\r (S_{\r+\yhat/2}
  - S_{\r-\yhat/2}) A^x_\r + (S_{\r-\xhat/2} - S_{\r+\xhat/2}) A^y_\r
  \\ & = & \sum_\r S_\r (A^x_{\r-\yhat/2} - A^x_{\r+\yhat/2} +
  A^y_{\r+\xhat/2} - A^y_{\r-\xhat/2}) \\ & = & \sum_\r S_\r (\D
  \times \A_\r) = 2\pi \sum_\r S_\r f_\r.
\end{eqnarray*}
\begin{displaymath}
  Z = \sum_{\{S_\r\}, \S} \exp\left(i2\pi \sum_\r S_\r f_\r \right)
  \prod_{\left<\r\r'\right>} \frac{e^{-[S_\r - S_{\r'}]^2/2\beta
      J}}{\sqrt{2\pi\beta J}}
\end{displaymath}
using
\begin{displaymath}
  \sum_S g(S) = \int d\sigma\sum_m e^{i2\pi m\sigma} g(\sigma),
\end{displaymath}
this becomes
\begin{displaymath}
  Z = \int \prod_{\r} d\sigma_\r d\vsigma
  \exp\left(-\sum_{\left<\r,\r'\right>}
  \frac{[\sigma_\r-\sigma_{\r'}]^2}{2\beta J} \right)
  \sum_{\{m_\r\},\m} \exp\left[i2\pi \left(\m \cdot \vsigma +
  \sum_{\r} (f_\r + m_\r)\sigma_\r\right)\right], 
\end{displaymath}
where $\m = (m_x, m_y)$.  After a Fourier transform this may be written
\begin{eqnarray}
  Z & = & \int \prod_{\k>0} (d\Re\sigma_\k)(d\Im\sigma_\k) d\vsigma
  \exp\left(\frac{1}{2\beta J L^2}\sum_\k\frac{|\sigma_\k|^2}{G(\k)} -
  \frac{1}{\beta J}\frac{\vsigma^2}{2} \right) \nonumber \\
  & & \sum_{\{m_\r\},\m} \exp\left(\frac{2\pi^2\beta J}{L^2} \sum_\k G(\k)
  |m_\k + f_\k|^2 - 2\pi^2\beta J \M^2 \right),
  \label{Z.mk}
\end{eqnarray}
where $\M = \m + \frac{1}{L} \sum_\r m_\r \r$ is the polarisation.
After a transformation back to ordinary space, this gives 
\begin{displaymath}
  Z = Z_{\rm sw} \sum_{\{m_\r\},\m} 
    \exp\left[-4\pi^2\beta J \left(-\frac{1}{2}\sum_{\r,\r'} (m_\r + f_\r)
    G(\r-\r') (m_{\r'} + f_{\r'}) + \frac{\M^2}{2} \right)\right]. 
\end{displaymath}
Dropping the 'spin wave part' $Z_{\rm sw}$, specializing to the frustration
$f_\r = 1/2$, and substituting $m_\r + f_\r \rightarrow m_\r$, we may then
write
\begin{equation}
  Z = Z_{\rm sw} \sum_{\{m_\r=\pm 1/2\ldots\},\m} \exp\left[-4\pi^2\beta J
  \left(-\frac{1}{2}\sum_{\r,\r'} m_\r G(\r-\r') m_{\r'} + \frac{\M^2}{2}
  \right)\right].
  \label{Zfinal}
\end{equation}
which is our final result for the partition function. 

We then turn to demonstrating the equivalence of Eqs.\ (\Ref{def.Upsk.m})
and (\Ref{def.Upsk.v}).  The wave-vector dependent helicity modulus is
defined through\cite{Ciordas_Teitel}
\begin{displaymath}
  \Upsilon(\k) = L^2 \frac{\partial^2 F}{\partial A^x_\k \partial
  A^x_{-\k}},
\end{displaymath}
where the derivative is with respect to an $\A_\k$ that is transverse, i.e.\ 
$\tk\;\cdot\;\A_\k = 0$.  Since only the rotation of $\A$ contributes to
the change in free energy, we make use of $2\pi f_\k = i\tk\times\A_\k$ to
obtain
\begin{displaymath}
  \Upsilon(\k) = \frac{\tk^2}{4\pi^2} L^2 \parti{^2 F}{f_\k \partial f_{-\k}}
\end{displaymath}
From Eq.\ (\Ref{Z.mk}) then follows
\begin{displaymath}
  \frac{1}{Z}\parti{^2Z}{f_\k \partial f_{-\k}}
  = \left<\frac{4\pi^2 J}{TL^2} G(\k) \right>
  + \left<\left(\frac{4\pi^2 J}{TL^2}\right)^2 |m_{-\k} + f_{-\k}|
  G(\k) |m_\k + f_\k| G(-\k) \right>,
\end{displaymath}
which, with $F = -T\ln Z$, $\tk^2 = -1/G(\k)$, and $m_\k + f_\k \rightarrow
m_\k$, reproduces Eq.\ (\Ref{def.Upsk.m}).

By instead starting from Eq.\ (\Ref{Zstart}) we find
\begin{displaymath}
  \frac{1}{Z}\parti{^2Z}{f_\k\partial f_{-\k}} = 
  \left<-\beta \sum_\r U''(\phi_\r) \left|\frac{d A_\r}{df_\k} \right|^2\right>
  + \left< \beta^2 \sum_\r U'(\phi_\r) \frac{d A_\r}{df_\k}
    \sum_\rp U'(\phi_\rp) \frac{d A_\rp}{df_{-\k}} \right>.
\end{displaymath}
To evaluate the derivatives we have to choose $\A_\r$ such that $\D\times
\A_\r = 2\pi f_\r$.  One choice that fulfills this is
\begin{displaymath}
  \A_\k = \frac{2\pi}{\tk^2} f_\k (i\tky,-i\tkx),
\end{displaymath}
that gives
\begin{displaymath}
  \frac{d \A_\r}{d f_\k} = \frac{2\pi}{\tk^2 L^2} (i\tky,-i\tkx)
  e^{i\k\cdot\r}.
\end{displaymath}
Putting all this together and introducing the current $\j_\r =
(U'(\phi^x_\r),U'(\phi^y_\r)$, the wave-vector dependent helicity modulus
becomes
\begin{displaymath}
  \Upsilon(\k) = L^2 \frac{\tk^2}{4\pi^2} \parti{^2 F}{f_\k \partial f_{-\k}}
  = \left<U''(\phi_\r)\right>
  - \frac{\beta}{L^2} \left<\j_\k \cdot \j_{-\k}\right>,
\end{displaymath}
and with $2\pi J v_\r = \D \times \j_\r\;\Leftrightarrow\; 2\pi J v_\k =
i\tk\times \j_\k$, we finally obtain Eq.\ (\Ref{def.Upsk.v}).

\newpage
\footnotesize 

\end{document}

%% file: gg.tex
\begin{picture}(87,60)(-13,-8)
\footnotesize
\xlabel{ 7.50}{0.78}
\xlabel{17.50}{0.80}
\xlabel{27.50}{0.82}
\xlabel{37.50}{0.84}
\xlabel{47.50}{0.86}
\xlabel{57.50}{0.88}
\xlabel{67.50}{0.90}
\ylabel{ 6.19}{0.244}
\ylabel{18.56}{0.246}
\ylabel{30.94}{0.248}
\ylabel{43.31}{0.250}
\small
\put(35,-6.53232){\makebox(0,0)[c]{$T/J$}}
\put(-10.5,23.8188){\rotate{\makebox(0,0)[c]{$\left<v^2\right>$,\hspace{1cm} $\left<m^2\right>$}}}

%% file: Ups.t.tex
\begin{picture}(87,60)(-13,-8)
\footnotesize
\xlabel{ 0.00}{0.810}
\xlabel{15.90}{0.815}
\xlabel{31.80}{0.820}
\xlabel{47.71}{0.825}
\xlabel{63.61}{0.830}
\ylabel{ 0.00}{0.40}
\ylabel{ 8.25}{0.45}
\ylabel{16.50}{0.50}
\ylabel{24.75}{0.55}
\ylabel{33.00}{0.60}
\ylabel{41.25}{0.65}
\ylabel{49.50}{0.70}
\small
\put(50.4,42.0085){\makebox(0,0)[l]{$L = 32$}}
\put(50.4,37.8076){\makebox(0,0)[l]{$L = 64$}}
\put(50.4,33.6068){\makebox(0,0)[l]{$L = 128$}}
\put(35,-6.53232){\makebox(0,0)[c]{$T/J$}}
\put(-10.5,23.8188){\rotate{\makebox(0,0)[c]{$\Upsilon/J$}}}

%% file: TKT.L.tex
\begin{picture}(87,60)(-13,-8)
\footnotesize
\xlabel{ 5.00}{0.00}
\xlabel{15.00}{0.01}
\xlabel{25.00}{0.02}
\xlabel{35.00}{0.03}
\xlabel{45.00}{0.04}
\xlabel{55.00}{0.05}
\xlabel{65.00}{0.06}
\ylabel{ 4.95}{0.81}
\ylabel{14.85}{0.82}
\ylabel{24.75}{0.83}
\ylabel{34.65}{0.84}
\ylabel{44.55}{0.85}
\small
\put(35,-6.53232){\makebox(0,0)[c]{$1/L$}}
\put(-10.5,23.8188){\rotate{\makebox(0,0)[c]{$T_{\rm KT}^{(L)}$}}}

%% file: m.L.tex
\begin{picture}(87,60)(-13,-8)
\footnotesize
\xlabel{ 0.00}{0.00}
\xlabel{10.00}{0.01}
\xlabel{20.00}{0.02}
\xlabel{30.00}{0.03}
\xlabel{40.00}{0.04}
\xlabel{50.00}{0.05}
\xlabel{60.00}{0.06}
\xlabel{70.00}{0.07}
\ylabel{ 4.13}{0.68}
\ylabel{12.38}{0.70}
\ylabel{20.62}{0.72}
\ylabel{28.87}{0.74}
\ylabel{37.12}{0.76}
\ylabel{45.38}{0.78}
\small
\put(35,-6.53232){\makebox(0,0)[c]{$1/L$}}
\put(-10.5,23.8188){\rotate{\makebox(0,0)[c]{$M_L(T_{\rm KT}^{(L)}$)}}}

%% file: m.q.tex
\begin{picture}(87,60)(-13,-8)
\footnotesize
\xlabel{ 0.00}{0.90}
\xlabel{17.50}{0.95}
\xlabel{35.00}{1.00}
\xlabel{52.50}{1.05}
\xlabel{70.00}{1.10}
\ylabel{ 0.00}{0.60}
\ylabel{ 5.50}{0.62}
\ylabel{11.00}{0.64}
\ylabel{16.50}{0.66}
\ylabel{22.00}{0.68}
\ylabel{27.50}{0.70}
\ylabel{33.00}{0.72}
\ylabel{38.50}{0.74}
\ylabel{44.00}{0.76}
\ylabel{49.50}{0.78}
\small
\put(53.9,4.20085){\makebox(0,0)[l]{$L = 16$}}
\put(53.9,8.4017){\makebox(0,0)[l]{$L = 32$}}
\put(53.9,12.6025){\makebox(0,0)[l]{$L = 48$}}
\put(53.9,16.8034){\makebox(0,0)[l]{$L = 64$}}
\put(53.9,21.0042){\makebox(0,0)[l]{$L = 96$}}
\put(53.9,25.2051){\makebox(0,0)[l]{$L = 128$}}
\put(35,-6.53232){\makebox(0,0)[c]{$\Upsilon\pi/2T$}}
\put(-10.5,23.8188){\rotate{\makebox(0,0)[c]{$M$}}}

%% file: Ups.Lmin.tex
\begin{picture}(87,60)(-13,-8)
\footnotesize
\xlabel{11.67}{16}
\xlabel{35.00}{32}
\xlabel{58.33}{64}
\ylabel{ 0.00}{0}
\ylabel{ 7.07}{1}
\ylabel{14.14}{2}
\ylabel{21.21}{3}
\ylabel{28.29}{4}
\ylabel{35.36}{5}
\ylabel{42.43}{6}
\ylabel{49.50}{7}
\small
\put(35,-6.53232){\makebox(0,0)[c]{$L_{\rm min}$}}
\put(-10.5,23.8188){\rotate{\makebox(0,0)[c]{$\chi^2/{\rm DOF}$}}}

%% file: Ups.L.tex
\begin{picture}(87,60)(-13,-8)
\footnotesize
\xlabel{ 9.21}{16}
\xlabel{27.63}{32}
\xlabel{46.05}{64}
\xlabel{64.47}{128}
\ylabel{ 0.00}{1.10}
\ylabel{ 7.73}{1.15}
\ylabel{15.47}{1.20}
\ylabel{23.20}{1.25}
\ylabel{30.94}{1.30}
\ylabel{38.67}{1.35}
\ylabel{46.41}{1.40}
\small
\put(36.4,35.0071){\makebox(0,0)[l]{$T_{\rm KT} = 0.8108$}}
\put(35,-6.53232){\makebox(0,0)[c]{$L$}}
\put(-10.5,23.8188){\rotate{\makebox(0,0)[c]{$\Upsilon_L\pi/(2T)$}}}

%% file: lo.c.sqrt.tex
\begin{picture}(87,60)(-13,-8)
\footnotesize
\xlabel{ 0.00}{0.00}
\xlabel{11.67}{0.02}
\xlabel{23.33}{0.04}
\xlabel{35.00}{0.06}
\xlabel{46.67}{0.08}
\xlabel{58.33}{0.10}
\xlabel{70.00}{0.12}
\ylabel{ 0.00}{0.00}
\ylabel{ 7.07}{0.05}
\ylabel{14.14}{0.10}
\ylabel{21.21}{0.15}
\ylabel{28.29}{0.20}
\ylabel{35.36}{0.25}
\ylabel{42.43}{0.30}
\ylabel{49.50}{0.35}
\small
\put(8.4,35.0071){\makebox(0,0)[l]{Slope: $B = 2.889$}}
\put(8.4,30.8062){\makebox(0,0)[l]{$\Rightarrow\; C = 0.544$}}
\put(8.4,42.0085){\makebox(0,0)[l]{(a)}}
\put(35,-6.53232){\makebox(0,0)[c]{$\sqrt{1 - T/T_{\rm KT}}$}}
\put(-10.5,23.8188){\rotate{\makebox(0,0)[c]{$c$}}}

%% file: hi.c.sqrt.tex
\begin{picture}(87,60)(-13,-8)
\footnotesize
\xlabel{ 0.00}{0.00}
\xlabel{ 9.46}{0.01}
\xlabel{18.92}{0.02}
\xlabel{28.38}{0.03}
\xlabel{37.84}{0.04}
\xlabel{47.30}{0.05}
\xlabel{56.76}{0.06}
\xlabel{66.22}{0.07}
\ylabel{ 0.00}{0.00}
\ylabel{11.25}{0.05}
\ylabel{22.50}{0.10}
\ylabel{33.75}{0.15}
\ylabel{45.00}{0.20}
\small
\put(8.4,35.0071){\makebox(0,0)[l]{Slope: $B = 2.834$}}
\put(8.4,30.8062){\makebox(0,0)[l]{$\Rightarrow\; C = 0.554$}}
\put(8.4,42.0085){\makebox(0,0)[l]{(b)}}
\put(35,-6.53232){\makebox(0,0)[c]{$\sqrt{T/T_{\rm KT} - 1}$}}
\put(-10.5,23.8188){\rotate{\makebox(0,0)[c]{$c$}}}

%% file: hi.l0.sqrt.tex
\begin{picture}(87,60)(-13,-8)
\footnotesize
\xlabel{ 2.69}{14}
\xlabel{13.46}{16}
\xlabel{24.23}{18}
\xlabel{35.00}{20}
\xlabel{45.77}{22}
\xlabel{56.54}{24}
\xlabel{67.31}{26}
\ylabel{ 0.00}{ 9}
\ylabel{ 7.07}{10}
\ylabel{14.14}{11}
\ylabel{21.21}{12}
\ylabel{28.29}{13}
\ylabel{35.36}{14}
\ylabel{42.43}{15}
\ylabel{49.50}{16}
\small
\put(8.4,42.0085){\makebox(0,0)[l]{Slope: $C = 0.532$}}
\put(35,-6.53232){\makebox(0,0)[c]{$1/\sqrt{T/T_{\rm KT} - 1}$}}
\put(-10.5,23.8188){\rotate{\makebox(0,0)[c]{$\ell_0$}}}

%% file: TKT.L.line.tex
\begin{picture}(87,60)(-13,-8)
\footnotesize
\xlabel{ 0.00}{ 4}
\xlabel{15.56}{ 6}
\xlabel{31.11}{ 8}
\xlabel{46.67}{10}
\xlabel{62.22}{12}
\ylabel{ 0.00}{2.5}
\ylabel{ 9.90}{3.0}
\ylabel{19.80}{3.5}
\ylabel{29.70}{4.0}
\ylabel{39.60}{4.5}
\ylabel{49.50}{5.0}
\small
\put(8.4,33.6068){\makebox(0,0)[l]{Slope: $C' = 0.265$}}
\put(8.4,40.6082){\makebox(0,0)[l]{(a)}}
\put(35,-6.53232){\makebox(0,0)[c]{$1/\sqrt{T_{\rm KT}^{(L)}/T_{\rm KT} - 1}$}}
\put(-10.5,23.8188){\rotate{\makebox(0,0)[c]{$\ln L$}}}

%% file: XY.TKT.L.tex
\begin{picture}(87,60)(-13,-8)
\footnotesize
\xlabel{ 7.78}{4}
\xlabel{23.33}{5}
\xlabel{38.89}{6}
\xlabel{54.44}{7}
\xlabel{70.00}{8}
\ylabel{ 2.47}{2}
\ylabel{14.85}{3}
\ylabel{27.23}{4}
\ylabel{39.60}{5}
\small
\put(8.4,35.0071){\makebox(0,0)[l]{Ordinary 2D $XY$}}
\put(8.4,30.8062){\makebox(0,0)[l]{Slope: $C' = 0.853$}}
\put(8.4,42.0085){\makebox(0,0)[l]{(b)}}
\put(35,-6.53232){\makebox(0,0)[c]{$1/\sqrt{T_{{\rm KT},XY}^{(L)}/T^{XY}_{\rm KT} - 1}$}}
\put(-10.5,23.8188){\rotate{\makebox(0,0)[c]{$\ln L$}}}

%% file: binder.tex
\begin{picture}(87,60)(-13,-8)
\footnotesize
\xlabel{ 0.00}{0.815}
\xlabel{17.50}{0.820}
\xlabel{35.00}{0.825}
\xlabel{52.50}{0.830}
\xlabel{70.00}{0.835}
\ylabel{ 0.00}{0.56}
\ylabel{ 9.90}{0.58}
\ylabel{19.80}{0.60}
\ylabel{29.70}{0.62}
\ylabel{39.60}{0.64}
\ylabel{49.50}{0.66}
\small
\put(11.9,21.0042){\makebox(0,0)[l]{$L = 8$}}
\put(11.9,16.8034){\makebox(0,0)[l]{$L = 16$}}
\put(11.9,12.6025){\makebox(0,0)[l]{$L = 32$}}
\put(11.9,8.4017){\makebox(0,0)[l]{$L = 64$}}
\put(35,42.0085){\makebox(0,0)[l]{(a)}}
\put(35,-6.53232){\makebox(0,0)[c]{$T/J$}}
\put(-10.5,23.8188){\rotate{\makebox(0,0)[c]{$U_L$}}}

%% file: binder.fit.tex
\begin{picture}(87,60)(-13,-8)
\footnotesize
\xlabel{ 7.00}{-0.4}
\xlabel{21.00}{-0.2}
\xlabel{35.00}{0.0}
\xlabel{49.00}{0.2}
\xlabel{63.00}{0.4}
\ylabel{ 0.00}{0.60}
\ylabel{11.00}{0.61}
\ylabel{22.00}{0.62}
\ylabel{33.00}{0.63}
\ylabel{44.00}{0.64}
\small
\put(8.4,18.2037){\makebox(0,0)[l]{$t = T/0.8244 - 1$}}
\put(50.4,37.8076){\makebox(0,0)[l]{$L = 16$}}
\put(50.4,33.6068){\makebox(0,0)[l]{$L = 32$}}
\put(50.4,29.4059){\makebox(0,0)[l]{$L = 64$}}
\put(8.4,14.0028){\makebox(0,0)[l]{$\nu = 0.808$}}
\put(35,42.0085){\makebox(0,0)[l]{(b)}}
\put(35,-6.53232){\makebox(0,0)[c]{$t L^{1/\nu}$}}
\put(-10.5,23.8188){\rotate{\makebox(0,0)[c]{$U_L$}}}

%% file: is.xi.xxi.tex
\begin{picture}(87,60)(-13,-8)
\footnotesize
\xlabel{ 0.00}{0.00}
\xlabel{10.77}{0.02}
\xlabel{21.54}{0.04}
\xlabel{32.31}{0.06}
\xlabel{43.08}{0.08}
\xlabel{53.85}{0.10}
\xlabel{64.62}{0.12}
\ylabel{ 4.50}{0.08}
\ylabel{13.50}{0.10}
\ylabel{22.50}{0.12}
\ylabel{31.50}{0.14}
\ylabel{40.50}{0.16}
\ylabel{49.50}{0.18}
\small
\put(35,-6.53232){\makebox(0,0)[c]{$1/r_{\rm min}$}}
\put(-10.5,23.8188){\rotate{\makebox(0,0)[c]{$1/\xi$}}}

%% file: is.kxi.t.tex
\begin{picture}(87,60)(-13,-8)
\footnotesize
\xlabel{ 9.35}{2.30}
\xlabel{24.51}{2.35}
\xlabel{39.67}{2.40}
\xlabel{54.84}{2.45}
\xlabel{70.00}{2.50}
\ylabel{ 0.00}{0.00}
\ylabel{ 5.50}{0.02}
\ylabel{11.00}{0.04}
\ylabel{16.50}{0.06}
\ylabel{22.00}{0.08}
\ylabel{27.50}{0.10}
\ylabel{33.00}{0.12}
\ylabel{38.50}{0.14}
\ylabel{44.00}{0.16}
\ylabel{49.50}{0.18}
\small
\put(8.4,36.4074){\makebox(0,0)[l]{$L = 64$}}
\put(8.4,32.2065){\makebox(0,0)[l]{$L = 128$}}
\put(8.4,28.0057){\makebox(0,0)[l]{$L = 256$}}
\put(35,-6.53232){\makebox(0,0)[c]{$T/J$}}
\put(-10.5,23.8188){\rotate{\makebox(0,0)[c]{$1/\xi$}}}

%% file: is.tc.tmax.tex
\begin{picture}(87,60)(-13,-8)
\footnotesize
\xlabel{ 8.96}{2.30}
\xlabel{23.49}{2.35}
\xlabel{38.03}{2.40}
\xlabel{52.56}{2.45}
\xlabel{67.09}{2.50}
\ylabel{ 0.00}{2.264}
\ylabel{ 9.90}{2.266}
\ylabel{19.80}{2.268}
\ylabel{29.70}{2.270}
\ylabel{39.60}{2.272}
\ylabel{49.50}{2.274}
\small
\put(8.4,9.80198){\makebox(0,0)[l]{$1/\xi \propto T - T_c$}}
\put(8.4,5.60113){\makebox(0,0)[l]{$1/\xi \propto (T - T_c)^\nu$}}
\put(8.4,42.0085){\makebox(0,0)[l]{(a)}}
\put(35,-6.53232){\makebox(0,0)[c]{$T_{\rm max}/J$}}
\put(-10.5,23.8188){\rotate{\makebox(0,0)[c]{$T_c^I/J$}}}

%% file: is.nu.tmax.tex
\begin{picture}(87,60)(-13,-8)
\footnotesize
\xlabel{ 8.96}{2.30}
\xlabel{23.49}{2.35}
\xlabel{38.03}{2.40}
\xlabel{52.56}{2.45}
\xlabel{67.09}{2.50}
\ylabel{ 0.00}{0.85}
\ylabel{ 9.90}{0.90}
\ylabel{19.80}{0.95}
\ylabel{29.70}{1.00}
\ylabel{39.60}{1.05}
\ylabel{49.50}{1.10}
\small
\put(8.4,42.0085){\makebox(0,0)[l]{(b)}}
\put(35,-6.53232){\makebox(0,0)[c]{$T_{\rm max}/J$}}
\put(-10.5,23.8188){\rotate{\makebox(0,0)[c]{$\nu$}}}

%% file: finite.64.tex
\begin{picture}(87,60)(-13,-8)
\footnotesize
\xlabel{ 0.00}{0.82}
\xlabel{11.11}{0.83}
\xlabel{22.22}{0.84}
\xlabel{33.33}{0.85}
\xlabel{44.44}{0.86}
\xlabel{55.56}{0.87}
\xlabel{66.67}{0.88}
\ylabel{ 0.00}{0.00}
\ylabel{12.38}{0.05}
\ylabel{24.75}{0.10}
\ylabel{37.13}{0.15}
\ylabel{49.50}{0.20}
\small
\put(50.4,36.4074){\makebox(0,0)[l]{$L = 64$}}
\put(50.4,32.2065){\makebox(0,0)[l]{$L = 128$}}
\put(35,42.0085){\makebox(0,0)[l]{(a)}}
\put(22.4,7.00141){\makebox(0,0)[l]{$\Upsilon^{(64)}$}}
\put(35,-6.53232){\makebox(0,0)[c]{$T/J$}}
\put(-10.5,23.8188){\rotate{\makebox(0,0)[c]{$g(q=0)/4000,\;\;\;\Upsilon/J$}}}

%% file: finite.128.tex
\begin{picture}(87,60)(-13,-8)
\footnotesize
\xlabel{ 0.00}{0.82}
\xlabel{11.11}{0.83}
\xlabel{22.22}{0.84}
\xlabel{33.33}{0.85}
\xlabel{44.44}{0.86}
\xlabel{55.56}{0.87}
\xlabel{66.67}{0.88}
\ylabel{ 0.00}{0.00}
\ylabel{ 9.90}{0.02}
\ylabel{19.80}{0.04}
\ylabel{29.70}{0.06}
\ylabel{39.60}{0.08}
\ylabel{49.50}{0.10}
\small
\put(50.4,32.2065){\makebox(0,0)[l]{$L = 128$}}
\put(50.4,28.0057){\makebox(0,0)[l]{$L = 256$}}
\put(35,42.0085){\makebox(0,0)[l]{(b)}}
\put(18.9,7.00141){\makebox(0,0)[l]{$\Upsilon^{(128)}$}}
\put(35,-6.53232){\makebox(0,0)[c]{$T/J$}}
\put(-10.5,23.8188){\rotate{\makebox(0,0)[c]{$g(q=2\pi/128)/4000,\;\;\;\Upsilon/J$}}}

%% file: epsk840.tex
\begin{picture}(87,60)(-13,-8)
\footnotesize
\xlabel{ 0.00}{0.0}
\xlabel{10.94}{0.5}
\xlabel{21.88}{1.0}
\xlabel{32.81}{1.5}
\xlabel{43.75}{2.0}
\xlabel{54.69}{2.5}
\xlabel{65.63}{3.0}
\ylabel{ 0.00}{0.0}
\ylabel{12.38}{0.2}
\ylabel{24.75}{0.4}
\ylabel{37.13}{0.6}
\ylabel{49.50}{0.8}
\small
\put(25.9,25.2051){\makebox(0,0)[l]{$L = 8$}}
\put(25.9,21.0042){\makebox(0,0)[l]{$L = 16$}}
\put(25.9,16.8034){\makebox(0,0)[l]{$L = 32$}}
\put(25.9,12.6025){\makebox(0,0)[l]{$L = 64$}}
\put(25.9,8.4017){\makebox(0,0)[l]{$L = 128$}}
\put(25.9,4.20085){\makebox(0,0)[l]{$L = 256$}}
\put(35,-6.53232){\makebox(0,0)[c]{$k$}}
\put(-10.5,23.8188){\rotate{\makebox(0,0)[c]{$\Upsilon(k)$}}}

%% file: kxi.Ups.t.tex
\begin{picture}(87,60)(-13,-8)
\footnotesize
\xlabel{ 0.00}{0.82}
\xlabel{11.29}{0.83}
\xlabel{22.58}{0.84}
\xlabel{33.87}{0.85}
\xlabel{45.16}{0.86}
\xlabel{56.45}{0.87}
\xlabel{67.74}{0.88}
\ylabel{ 0.00}{0.00}
\ylabel{ 5.50}{0.02}
\ylabel{11.00}{0.04}
\ylabel{16.50}{0.06}
\ylabel{22.00}{0.08}
\ylabel{27.50}{0.10}
\ylabel{33.00}{0.12}
\ylabel{38.50}{0.14}
\ylabel{44.00}{0.16}
\ylabel{49.50}{0.18}
\small
\put(50.4,22.4045){\makebox(0,0)[l]{$L = 64$}}
\put(50.4,18.2037){\makebox(0,0)[l]{$L = 128$}}
\put(50.4,14.0028){\makebox(0,0)[l]{$L = 256$}}
\put(8.4,42.0085){\makebox(0,0)[l]{(a)}}
\put(21,35.0071){\makebox(0,0)[l]{$\Upsilon^{(64)}$}}
\put(4.9,28.0057){\makebox(0,0)[l]{$\Upsilon^{(128)}$}}
\put(35,-6.53232){\makebox(0,0)[c]{$T/J$}}
\put(-10.5,23.8188){\rotate{\makebox(0,0)[c]{$1/\xi$}}}

%% file: kxi.t.tex
\begin{picture}(87,60)(-13,-8)
\footnotesize
\xlabel{ 0.00}{0.82}
\xlabel{13.73}{0.83}
\xlabel{27.45}{0.84}
\xlabel{41.18}{0.85}
\xlabel{54.90}{0.86}
\xlabel{68.63}{0.87}
\ylabel{ 0.00}{0.00}
\ylabel{ 7.07}{0.02}
\ylabel{14.14}{0.04}
\ylabel{21.21}{0.06}
\ylabel{28.29}{0.08}
\ylabel{35.36}{0.10}
\ylabel{42.43}{0.12}
\ylabel{49.50}{0.14}
\small
\put(50.4,18.2037){\makebox(0,0)[l]{$L = 128$}}
\put(50.4,14.0028){\makebox(0,0)[l]{$L = 256$}}
\put(8.4,42.0085){\makebox(0,0)[l]{(b)}}
\put(35,-6.53232){\makebox(0,0)[c]{$T/J$}}
\put(-10.5,23.8188){\rotate{\makebox(0,0)[c]{$1/\xi$}}}

%% file: tc.tmax.tex
\begin{picture}(87,60)(-13,-8)
\footnotesize
\xlabel{ 0.00}{0.82}
\xlabel{13.73}{0.83}
\xlabel{27.45}{0.84}
\xlabel{41.18}{0.85}
\xlabel{54.90}{0.86}
\xlabel{68.63}{0.87}
\ylabel{ 0.00}{0.820}
\ylabel{ 9.90}{0.821}
\ylabel{19.80}{0.822}
\ylabel{29.70}{0.823}
\ylabel{39.60}{0.824}
\small
\put(8.4,9.80198){\makebox(0,0)[l]{$1/\xi \propto T - T_c$}}
\put(8.4,5.60113){\makebox(0,0)[l]{$1/\xi \propto (T - T_c)^\nu$}}
\put(8.4,42.0085){\makebox(0,0)[l]{(a)}}
\put(35,-6.53232){\makebox(0,0)[c]{$T_{\rm max}/J$}}
\put(-10.5,23.8188){\rotate{\makebox(0,0)[c]{$T_c/J$}}}

%% file: nu.tmax.tex
\begin{picture}(87,60)(-13,-8)
\footnotesize
\xlabel{ 0.00}{0.82}
\xlabel{13.73}{0.83}
\xlabel{27.45}{0.84}
\xlabel{41.18}{0.85}
\xlabel{54.90}{0.86}
\xlabel{68.63}{0.87}
\ylabel{ 0.00}{0.80}
\ylabel{ 8.25}{0.85}
\ylabel{16.50}{0.90}
\ylabel{24.75}{0.95}
\ylabel{33.00}{1.00}
\ylabel{41.25}{1.05}
\ylabel{49.50}{1.10}
\small
\put(8.4,42.0085){\makebox(0,0)[l]{(b)}}
\put(35,-6.53232){\makebox(0,0)[c]{$T_{\rm max}/J$}}
\put(-10.5,23.8188){\rotate{\makebox(0,0)[c]{$\nu$}}}

%% file: lambda.tex
\begin{picture}(87,60)(-13,-8)
\footnotesize
\xlabel{ 0.00}{0.82}
\xlabel{11.29}{0.83}
\xlabel{22.58}{0.84}
\xlabel{33.87}{0.85}
\xlabel{45.16}{0.86}
\xlabel{56.45}{0.87}
\xlabel{67.74}{0.88}
\ylabel{ 0.00}{0.0}
\ylabel{ 9.90}{0.1}
\ylabel{19.80}{0.2}
\ylabel{29.70}{0.3}
\ylabel{39.60}{0.4}
\ylabel{49.50}{0.5}
\small
\put(50.4,22.4045){\makebox(0,0)[l]{$L = 64$}}
\put(50.4,18.2037){\makebox(0,0)[l]{$L = 128$}}
\put(50.4,14.0028){\makebox(0,0)[l]{$L = 256$}}
\put(8.4,42.0085){\makebox(0,0)[l]{(a)}}
\put(35,-6.53232){\makebox(0,0)[c]{$T/J$}}
\put(-10.5,23.8188){\rotate{\makebox(0,0)[c]{$1/\lambda$}}}

%% file: ln.lambda.tex
\begin{picture}(87,60)(-13,-8)
\footnotesize
\xlabel{ 3.78}{3.5}
\xlabel{13.24}{4.0}
\xlabel{22.70}{4.5}
\xlabel{32.16}{5.0}
\xlabel{41.62}{5.5}
\xlabel{51.08}{6.0}
\xlabel{60.54}{6.5}
\xlabel{70.00}{7.0}
\ylabel{ 0.00}{0.8}
\ylabel{ 6.19}{1.0}
\ylabel{12.38}{1.2}
\ylabel{18.56}{1.4}
\ylabel{24.75}{1.6}
\ylabel{30.94}{1.8}
\ylabel{37.13}{2.0}
\ylabel{43.31}{2.2}
\small
\put(8.4,35.0071){\makebox(0,0)[l]{Slope: $C = 0.421$}}
\put(50.4,18.2037){\makebox(0,0)[l]{$L = 128$}}
\put(50.4,14.0028){\makebox(0,0)[l]{$L = 256$}}
\put(8.4,42.0085){\makebox(0,0)[l]{(b)}}
\put(35,-6.53232){\makebox(0,0)[c]{$1/\sqrt{T/T_{\rm KT} - 1}$}}
\put(-10.5,23.8188){\rotate{\makebox(0,0)[c]{$\ln\lambda$}}}

%% file: tw.810.tex
\begin{picture}(87,60)(-13,-8)
\footnotesize
\xlabel{ 0.00}{ 0}
\xlabel{10.77}{10}
\xlabel{21.54}{20}
\xlabel{32.31}{30}
\xlabel{43.08}{40}
\xlabel{53.85}{50}
\xlabel{64.62}{60}
\ylabel{ 0.00}{0.58}
\ylabel{ 7.07}{0.60}
\ylabel{14.14}{0.62}
\ylabel{21.21}{0.64}
\ylabel{28.29}{0.66}
\ylabel{35.36}{0.68}
\ylabel{42.43}{0.70}
\ylabel{49.50}{0.72}
\small
\put(50.4,40.6082){\makebox(0,0)[l]{$L = 32$}}
\put(50.4,36.4074){\makebox(0,0)[l]{$L = 64$}}
\put(50.4,32.2065){\makebox(0,0)[l]{$L = 128$}}
\put(29.4,28.0057){\makebox(0,0)[l]{PBC}}
\put(29.4,4.90099){\makebox(0,0)[l]{FBC}}
\put(15.4,42.0085){\makebox(0,0)[l]{a)}}
\put(35,-6.53232){\makebox(0,0)[c]{$r$}}
\put(-10.5,23.8188){\rotate{\makebox(0,0)[c]{$g$}}}

%% file: tw.820.tex
\begin{picture}(87,60)(-13,-8)
\footnotesize
\xlabel{ 0.00}{ 0}
\xlabel{10.77}{10}
\xlabel{21.54}{20}
\xlabel{32.31}{30}
\xlabel{43.08}{40}
\xlabel{53.85}{50}
\xlabel{64.62}{60}
\ylabel{ 3.30}{0.46}
\ylabel{ 9.90}{0.48}
\ylabel{16.50}{0.50}
\ylabel{23.10}{0.52}
\ylabel{29.70}{0.54}
\ylabel{36.30}{0.56}
\ylabel{42.90}{0.58}
\ylabel{49.50}{0.60}
\small
\put(50.4,40.6082){\makebox(0,0)[l]{$L = 32$}}
\put(50.4,36.4074){\makebox(0,0)[l]{$L = 64$}}
\put(50.4,32.2065){\makebox(0,0)[l]{$L = 128$}}
\put(29.4,42.0085){\makebox(0,0)[l]{b)}}
\put(35,-6.53232){\makebox(0,0)[c]{$r$}}
\put(-10.5,23.8188){\rotate{\makebox(0,0)[c]{$g$}}}

%% file: tw.M.t.tex
\begin{picture}(87,60)(-13,-8)
\footnotesize
\xlabel{ 5.38}{0.800}
\xlabel{16.15}{0.802}
\xlabel{26.92}{0.804}
\xlabel{37.69}{0.806}
\xlabel{48.46}{0.808}
\xlabel{59.23}{0.810}
\xlabel{70.00}{0.812}
\ylabel{ 0.00}{0.77}
\ylabel{ 7.07}{0.78}
\ylabel{14.14}{0.79}
\ylabel{21.21}{0.80}
\ylabel{28.29}{0.81}
\ylabel{35.36}{0.82}
\ylabel{42.43}{0.83}
\ylabel{49.50}{0.84}
\small
\put(35,-6.53232){\makebox(0,0)[c]{$T/J$}}
\put(-10.5,23.8188){\rotate{\makebox(0,0)[c]{$M$}}}

%% file: tw.xi.tlow.tex
\begin{picture}(87,60)(-13,-8)
\footnotesize
\xlabel{ 2.12}{0.77}
\xlabel{12.73}{0.78}
\xlabel{23.33}{0.79}
\xlabel{33.94}{0.80}
\xlabel{44.55}{0.81}
\xlabel{55.15}{0.82}
\xlabel{65.76}{0.83}
\ylabel{ 0.00}{0.0}
\ylabel{11.79}{0.1}
\ylabel{23.57}{0.2}
\ylabel{35.36}{0.3}
\ylabel{47.14}{0.4}
\small
\put(35,-6.53232){\makebox(0,0)[c]{$T/J$}}
\put(-10.5,23.8188){\rotate{\makebox(0,0)[c]{$1/\xi$}}}

%% file: p3.bbl
\begin{thebibliography}{10}
\bibitem{C-values}
Some values for this constant may be found in Table IV in Ref.\
  \cite{Olsson:Kost-fit}.
\end{thebibliography}
